\documentclass[fleqn,usenatbib]{mnras}

\usepackage{newtxtext,newtxmath}
\usepackage[T1]{fontenc}

\DeclareRobustCommand{\VAN}[3]{#2}
\let\VANthebibliography\thebibliography
\def\thebibliography{\DeclareRobustCommand{\VAN}[3]{##3}\VANthebibliography}

\usepackage{graphicx}	
\usepackage{amsmath}	

\title[Dynamical Properties of Globular Clusters]{
Investigating Dynamical Properties of Globular Clusters through a Family of Lowered Isothermal Models}

\author[Cheng and Jiang]{
Chia-Hsuan Cheng$^{1}$
and Ing-Guey Jiang$^{1,2}$
\\
$^{1}$Department of Physics, National Tsing-Hua University, Hsinchu, Taiwan\\
$^{2}$Institute of Astronomy, National Tsing-Hua University, Hsinchu, Taiwan\\
}

\date{Accepted XXX. Received YYY; in original form ZZZ}

\pubyear{20XX}

\begin{document}
\label{firstpage}
\pagerange{\pageref{firstpage}--\pageref{lastpage}}
\maketitle

\begin{abstract}
To investigate the dynamical properties of globular clusters, 
the surface brightness and kinematic data were collected and fitted to a family of lowered isothermal models called \textsc{\scriptsize LIMEPY} models. 
For 18 studied globular clusters, the amounts of concentration, 
truncation, and anisotropy were determined. 
In addition, the cluster mass, half-mass radius, distance, and mass-to-light ratio were also obtained.  
In general, \textsc{\scriptsize LIMEPY} models could describe these clusters well. 
Among these 18 clusters,  NGC 5139,  NGC 6388, and NGC 7078 were claimed to be candidates to host 
intermediate-mass black holes in literature.  
The models could not appropriately fit the central proper-motion velocity dispersion of NGC 5139 and the slope of proper-motion velocity-dispersion profile of NGC 6388. 
Thus, more dedicated models with intermediate-mass black holes or a group of stellar-mass black holes at cluster centers may need to be considered.
Considering NGC 7078, our model with some degree of anisotropy can fit the data. 
Finally, the strong concentration-truncation anti-correlation and truncation-semimajor-axis correlation 
were revealed, which could be the observational imprint
of the dynamical evolution of globular clusters.  
\end{abstract}


\section{Introduction}

Globular clusters are one of the oldest objects in the universe \citep{Vandenberg1996}.
They extend spherically in several or tens of parsecs with hundreds of thousands of stars \citep{Harris1996}.
The high stellar densities make them the primary venue for hosting exotic objects like millisecond pulsars \citep{Manchester1991} and blue stragglers \citep{Bailyn1995}. 
Globular clusters have been proposed to possibly also host intermediate-mass black holes \citep{Ebisuzaki2001}.
With higher density, the core of a globular cluster relaxes faster than the halo
and the relaxation time is short compared to the age of the cluster \citep{Oort1959}.
Thus, the center of globular clusters is expected to be isothermal.

Having theoretical models describing globular clusters is helpful in obtaining the physical quantities.
The isothermal sphere is a model with isothermal cores,
so it could be considered a suitable simple model.
However, this model extends to the infinite and has an unrealistic infinite mass. 
This problem can be solved by introducing some cutoffs.
For example, energy truncation can limit the velocity,
so the stars with larger velocities escape from the cluster; this results in a cluster model with finite mass and range.
The truncation can be regarded as the effect of the external tidal field on star clusters.
Different truncations lead to different models.
For example, subtracting a constant from the energy leads to the Woolley model \citep{Woolley1954},
and further subtraction from the distribution function gives the King model \citep{King1966}.

The velocity distributions of clusters in the above models are isotropic.  
However, for realistic models, the possible anisotropy shall be considered.
The diffusion caused by stellar encounters facilitates the entry of some stars into the cluster halo.
These stars diffuse to the halo along radial orbits and increase the radial anisotropy in the halo \citep{Spitzer1972}.
The violent relaxation in the stage of cluster formation can also contribute to some radial anisotropy in the cluster halo \citep{Lynden-Bell1967}.
To include anisotropy in a model, one can add the angular momentum into the distribution function.
The distribution function now depends on both the energy and the angular momentum.
For example, the Michie-King model \citep{Michie1963} includes the angular momentum in an exponential term.
This model possesses the expected properties which contain an isothermal core with some anisotropy 
at the outer parts.

A model with multi-mass components is another aspect of improvement. 
\citet{DaCosta1976} made the extension from the King model by assuming that each component has the same form of distribution function with different constants. 
Later, an anisotropic multi-mass model was introduced by \citet{Gunn1979}.
Recently, some extensions and unification of these isothermal models have been developed.
Considering the Woolley and the King model as different schemes of energy truncation characterized by some integers, \citet{Gomez-Leyton2014} established an extended model which parametrized the truncation by a non-negative real number.
This was further generalized by \citet{Gieles2015} to include the radial anisotropy and multi-mass components in a family of lowered isothermal models, which can cover more properties of star clusters. 
They also provided a fast model solver written as a Python code, \textsc{\scriptsize LIMEPY}, for this family of lowered isothermal models. 
Thus, these models proposed by \citet{Gieles2015} are called \textsc{\scriptsize LIMEPY} models.

As presented by \citet{Zocchi2016}, \textsc{\scriptsize LIMEPY} models could capture the main properties of the globular clusters. 
Moreover, \citet{Zocchi2017} applied \textsc{\scriptsize LIMEPY} models in the study of NGC 5139
and found that part of the observed large central velocity dispersion could be produced by anisotropic models.
Thus, their results could provide some constraints on the previously proposed central intermediate-mass black hole in NGC 5139 
\citep{Noyola2010}.
This globular cluster, also named $\omega$ Centauri, is the most complex
one which has many sub-populations \citep{Sanna+2020} and was heavily investigated with many 
controversial results. 
On the other hand, the central kinematics of NGC 6093 was studied by employing new integral-field spectrograph data, and the existence of an intermediate-mass black hole was supported \citep{Gottgens+2021}. 
In addition, NGC 6388 is also a candidate residence of the intermediate-mass black hole
\citep{Lutzgendorf2011}.

Moreover, with Gaia data, \citet{Vasiliev2021} performed a comprehensive study on the kinematic properties of many Galactic globular clusters. 
The proper motions were measured and the corresponding proper-motion dispersion profiles of 100 clusters were obtained. 
Combining with HST and other literature data, \citet{Baumgardt2021} also accurately derived the distances to these Galactic globular clusters.

Therefore, motivated by the development of \textsc{\scriptsize LIMEPY} models, the controversial results of the central kinematics and intermediate-mass black holes, and the availability of new data derived from the Gaia mission, herein, we investigated the properties of 18 globular clusters 
with the \textsc{\scriptsize LIMEPY} models.
Including data from recent observations such as the MUSE survey \citep{Kamann2018} and Gaia mission \citep{Vasiliev2021}, the physical parameters of these clusters were obtained through the data-model fitting.
Our results could lead to updated and accurate descriptions of the dynamical states of these clusters for the cases in which the data could be well fitted by the \textsc{\scriptsize LIMEPY} models which can be isotropic or anisotropic. 
Our results might also imply the possible existence of intermediate-mass black holes
for some globular clusters.

For the rest of this paper, in Section 2, we introduce the model’s distribution function and essential properties.
The observational data are described in Section 3, and the parameter determination method is shown in Section 4.
The results and discussions are presented in Section 5.
In Section 6, some conclusions are made.

\section{The Model}

The \textsc{\scriptsize LIMEPY} models were employed as the standard model in this study. 
As presented \citep{Gieles2015}, there are single-mass and multi-mass cases in \textsc{\scriptsize LIMEPY} models.
Considering the single-mass models, the distribution functions have the following form:
\begin{equation}
f(E,J)=A\exp\left(\frac{-J^2}{2r^2_{\text{a}}s^2}\right)E_{\gamma}\left(g,\frac{\phi(r_{\text{t}}) -E}{s^2}\right),
\end{equation}
for $E\leq \phi(r_{\text{t}})$ and $f(E,J)=0$ for $E>\phi(r_{\text{t}})$.
The function $E_\gamma(g,x)$ represents $e^x$ for $g=0$ and $e^x \gamma(g,x)/\Gamma(g)$ for $g>0$,
where $\gamma(g,x)$ is the lower incomplete gamma function and $\Gamma(g)$ stands for the gamma function.
This distribution function depends on the specific energy $E$ and the specific angular momentum $J$.
The function $\phi$ is the gravitational potential and $r_{\text{t}}$ is the truncation radius.
The parameter $g$ is called the truncation parameter, and it regulates the energy truncation of the model.
The parameter $r_{\text{a}}$ is the anisotropic radius, and it determines how anisotropic a system is.
When $r_{\text{a}}$ grows, the model is less anisotropic, and $r_{\text{a}} \rightarrow \infty$ corresponds to an isotropic model.
The constants $A$ and $s$ are used to set the physical scale of the model.
The density can be obtained by integrating the distribution function $f(E,J)$ over the velocity space:
\begin{equation}
\rho=\int f(E,J)\: \mbox{d}^3v.
\end{equation}
Since $E=v^2/2+\phi(r)$ and the distribution function is zero for $E > \phi(r_{\text{t}})$, it can be just integrated from 0 to $v_{\text{max}}=[2\phi(r_{\text{t}})-2\phi(r)]^{1/2}$ at each $r$.
This $v_{\text{max}}$ becomes zero when $r=r_{\text{t}}$ and the density vanishes for $r \geq r_{\text{t}}$.
Hence, the truncation radius $r_{\text{t}}$ represents the distance where the density comes to zero.

The gravitational potential $\phi$ is subjected to the Poisson equation. 
For spherical systems such as globular clusters, the equation results in the following form:
\begin{equation}
\frac{\mbox{d}^2\phi}{\mbox{d}r^2}+\frac{2}{r}\frac{\mbox{d}\phi}{\mbox{d}r}=4\pi G\rho,
\end{equation}
where $r$ is the radial coordinate and $G$ is the gravitational constant.
The relevant quantities were first turned into dimensionless ones for solving the Poisson equation.
The dimensionless potential is defined as $\hat{\phi}=[\phi(r_{\text{t}})-\phi]/s^2$.
The dimensionless density and radius are $\hat{\rho}=\rho/\rho_0$ and $\hat{r}=r/r_0$, where $\rho_0$ and $r_0$ satisfy $4\pi G r_0^2\rho_0/s^2=9$.
Then, the Poisson equation becomes
\begin{equation}
\frac{\mbox{d}^2\hat{\phi}}{\mbox{d}\hat{r}^2}+\frac{2}{\hat{r}}\frac{\mbox{d}\hat{\phi}}{\mbox{d}\hat{r}}=-9\hat{\rho}.
\end{equation}
The equation is solved with the boundary conditions that, at $\hat{r}=0$, 
$\mbox{d}\hat{\phi}/\mbox{d}\hat{r}=0$ and $\hat{\phi}=W_0$, where $W_0$ is a constant that specifies a particular solution.
Hence, $W_0$ is also a parameter of the \textsc{\scriptsize LIMEPY} model, called the concentration parameter.
It characterizes the concentration of the model.

As previously mentioned, \textsc{\scriptsize LIMEPY} models provide an extended family of isothermal models.
Those famous models are included as sub-families.
For example, the Woolley model \citep{Woolley1954} can be produced by setting $g=0, r_{\text{a}} \rightarrow \infty$.
When $g=1$ and $r_{\text{a}} \rightarrow \infty$, the King model \citep{King1966} is obtained.
The Wilson model \citep{Wilson1975}, which is more extended, corresponds to $g=2$ and $r_{\text{a}} \rightarrow \infty$.
Models with $W_0 \rightarrow \infty$ or $g \rightarrow \infty$ become the isothermal spheres.
In addition, the polytrope can be represented as $W_0 \rightarrow 0$.
It includes the Plummer model \citep{Plummer1911} which corresponds to the model with $g=3.5$.
It has a finite mass but infinite extents.
In general, the model with appropriate $W_0$ and $r_{\text{a}}$ can be finite in extent if $g<3.5$ and conversely infinite in extent with $g \geq 3.5$.
In addition, \citet{Gieles2015} also showed that one kind of finite model is unsuitable for star clusters.
These systems have an upturn in the density far from the center, so there is a large amount of mass in the halo.
The ratio of the virial radius and half-mass radius $r_{\text{v}}/r_{\text{h}}$ is a crucial parameter for these models.
They suggested that the models with $r_{\text{v}}/r_{\text{h}} \geq 0.64$ can adequately describe star clusters.

The \textsc{\scriptsize LIMEPY} models describe spherical systems with different concentrations, truncation, and radial anisotropy.
In general, the model is isotropic near the center but could be anisotropic in the middle part of the system.
The energy truncation limits the contribution of anisotropy to radial orbits with $E \approx \phi(r_{\text{t}})$ and thus suppresses the degree of radial anisotropy near the edge.
The corresponding physical picture is that a cluster under the interaction of an external tidal field has a preferential mass loss on stars with radial orbits.
This reduces the amount of anisotropy in the outer region \citep{Oh1992, Takahashi1997}.
Simulations of star clusters in the tidal field confirmed this isotropic behavior near the edge \citep{Tiongco2016}.
Thus, the energy truncation acts as a role of the tidal field.
In fact, the tidal field can also make the outer region profiles tangentially anisotropic \citep{Baumgardt2003}.

In addition to the anisotropic radius $r_{\text{a}}$, there is a convenient anisotropic parameter $\kappa\equiv 2K_{\text{r}}/K_{\text{t}}$, where $K_{\text{r}}$ is the total radial kinetic energy and $K_{\text{t}}$ is the total tangential kinetic energy.
If $\kappa>1$, the system is radially anisotropic, and if $\kappa<1$, the system is tangentially anisotropic.
When $\kappa=1$, it is an isotropic system.
Therefore, $\kappa$ represents a simple and global measure of the anisotropy.
We mainly used $\kappa$ to determine the amount of the anisotropy of clusters.

In \citet{Zocchi2016}, the comparisons with N-body simulations illustrated the variation of model parameters of a cluster during the evolution. 
The cluster started with the Plummer model and the simulation snapshots at different time were fitted with \textsc{\scriptsize LIMEPY} models. 
The concentration parameter tended to increase with time, which was also suggested previously by \citet{King1966}.
The truncation parameter $g$ decreased roughly from 2.5 to 0.5 during the evolution.
It corresponded to an increased truncation by the tidal field as a cluster gradually filled the Roche volume.
Thus, a cluster tends to become more concentrated and truncated with time.
In addition, the degree of radial anisotropy increased due to radial diffusion but decreased later during the core collapse.

\section{The Observational Data}

One of our primary goals is to provide updated results with a complete inclusion of all available observational data for globular clusters. 
The observational data of $V$-band surface brightness $\mu$ were taken from \citet{Trager1995}, 
which provided a catalog of surface brightness profiles for over a hundred Galactic globular clusters.
Some procedures were needed before the data were ready for the fitting.
There was a correction related to extinction.
The method is based on the global mean curve discussed in \citet{Fitzpatrick1999}, which uses the mean value for the ratio of the extinction $A_V$ and the reddening $E(B-V)$ so that $A_V = 3.1 E(B-V)$.
We took the reddening in the catalog of \citet{Harris1996} (2010 version) and then computed the corrected surface brightness by $\mu_i=\mu_{i,0}-A_V$, where $\mu_{i,0}$ denotes the data before the correction.
The data with $w_i<0.15$ were not adopted according to \citet{McLaughlin2005}, where $w_i$ is the weight of each data given in \citet{Trager1995}.

Because the data number was large, which might make the surface brightness dominate the fitting, we sliced the radial range with equal logarithmic width and averaged the surface brightness and the weight in each bin.
The bin number was 55 which equaled the largest data number of the velocity dispersion.
To compute the uncertainty for each data, we followed the method in \citet{McLaughlin2005}.
The uncertainty of the data was obtained by $\epsilon_{\mu,i}=\epsilon_{\mu,\text{b}}/w_i$, where $\epsilon_{\mu,\text{b}}$ is the base error bar for each cluster.



For line-of-sight velocity dispersion, we used the profiles derived from the collected literature \citep{Baumgardt2017}, the data from unpublished spectra of stars in the ESO and Keck Science archives \citep{Baumgardt2018}, and the dispersion from the integral-field-unit data from the WAGGS project \citep{Dalgleish2020}. 
The above data are expressed by open circles in Fig.~\ref{fig:Vlm-r}.
The data from the MUSE survey \citep{Kamann2018} were also used and denoted by solid triangles.
Some additional data were supplemented and marked as crosses, such as those from \citet{McLaughlin2006} for NGC 104 and Larson \& Seth (2015, private communication) for NGC 1851 and NGC 2808.
(The data of \citet{McLaughlin2006} and Larson \& Seth (2015, private communication) were collected from the compilation in \citet{Watkins2015b} and others were collected from the compilation in the updated web catalog (third version) of \citet{Baumgardt2018}.)

For proper-motion velocity dispersion, we mainly took the data from the Hubble Space Telescope from \citet{Watkins2015a} and the Gaia data from \citet{Vasiliev2021}.
Open circles expressed the former, and solid triangles expressed the latter in Fig.~\ref{fig:Vpm-r}.
Some additional data were supplemented and denoted by crosses, which include \citet{Haberle2021} for NGC 6441, \citet{McLaughlin2006} for NGC 104, \citet{McNamara2003} for NGC 7078, \citet{McNamara2012} for NGC 6266, and \citet{Zloczewski2012} for NGC 6656 and NGC 6752.
(The data of \citet{Vasiliev2021} and \citet{Haberle2021} were collected from the updated web catalog of \citet{Baumgardt2018}, and the data of \citet{McLaughlin2006}, \citet{McNamara2003}, \citet{McNamara2012}, and \citet{Zloczewski2012} were collected from \citet{Watkins2015b}.)



Some proper motion data were downloaded in units of km/s, which depends on the cluster distance written in the literature.
These data were transformed into mas/yr as the observational values for our work here.
The transformation is $v=v_0/DC$, where $v$ and $v_0$ are the velocity in mas/yr and km/s, $D$ is the distance and $C=4.74047 \;\text{km}\:\text{yr}\:\text{kpc}^{-1}\:\text{mas}^{-1}\:\text{s}^{-1}$ which is a factor for the unit conversion \citep{vanLeeuwen2009, Watkins2015b}.
The values of cluster distances were taken from the corresponding literature.
By taking the root mean square of the upper and lower error bars from the literature, we obtained 
a symmetric uncertainty for each data for our work.
Finally, to focus on the systems with enough observational information, 
we studied 18 clusters with more than five data points in each type of the above observational profiles.

\section{The Determination of Physical Parameters}

It was shown in \citet{Zocchi2017} that models with different amounts of anisotropy could give the same surface brightness but different kinematic profiles. 
Thus, using the surface brightness data alone can lead to some degeneracy.
Therefore, here we included the surface brightness, the light-of-sight velocity dispersion, 
and the proper-motion velocity dispersion data to obtain complete pictures of the  
physical structures and kinematic properties of globular clusters by determining related physical parameters through 
the data-model fitting. 

Following the method in \citet{Zocchi2017}, we employed the one-step fitting procedure with the single-mass \textsc{\scriptsize LIMEPY} models in this paper.
With all three considered types of observational data, a single step of the fitting was performed to determine all cluster parameters. 
The fitting was done through the minimization of the $\chi^2$ function:
\begin{equation}
\chi^2 = \chi^2_{\text{sb}} + \chi^2_{\text{los}} + \chi^2_{\text{pm}},
\end{equation}
where $\chi^2_{\text{sb}}$, $\chi^2_{\text{los}}$, $\chi^2_{\text{pm}}$
are the contributions from surface brightness, line-of-sight velocity dispersion, and proper-motion velocity dispersion,
respectively. 
They are defined by
\begin{equation}
\chi^2_{\text{sb}} = \sum^{n_\text{{sb}}}_{i=1} \frac{ [\mu_i-\bar{\mu}(r_i)]^2 }{\epsilon_{\mu,i}^2},
\end{equation}
\begin{equation}
\chi^2_{\text{los}} = \sum^{n_\text{{los}}}_{i=1} \frac{ [\sigma_{\text{los},i}-\bar{\sigma}_\text{los}(r_i)]^2 }{\epsilon_{\text{los},i}^2},
\end{equation}
and
\begin{equation}
\chi^2_{\text{pm}} = \sum^{n_\text{{pm}}}_{i=1} \frac{ [\sigma_{\text{pm},i}-\bar{\sigma}_\text{pm}(r_i)]^2 }{\epsilon_{\text{pm},i}^2},
\end{equation}
where $\mu_i$ is the $i$-th observational data of a surface brightness profile, $\bar{\mu}(r_i)$
is the theoretical surface brightness at that radial coordinate $r_i$, and $\epsilon_{\mu,i}$ is the
error bar of the data $\mu_i$.
Similarly, $\sigma_{\text{los},i}$,  $\bar{\sigma}_\text{los}(r_i)$, $\epsilon_{\text{los},i}$
are the corresponding quantities for line-of-sight velocity dispersion,
and $\sigma_{\text{pm},i}$,  $\bar{\sigma}_\text{pm}(r_i)$, $\epsilon_{\text{pm},i}$
are the observational data, theoretical value, and error bar for proper-motion velocity dispersion,
respectively. 
The numbers of observational data are $n_\text{{sb}}$, $n_\text{{los}}$, $n_\text{{pm}}$, respectively, for the surface brightness, line-of-sight velocity dispersion, and proper-motion velocity dispersion, individually.

The \textsc{\scriptsize LIMEPY} code was employed to obtain the above theoretical profiles. 
This code needed five input parameters, including the concentration parameter $W_0$, the truncation parameter $g$, 
the dimensionless anisotropy radius $\hat{r}_\text{a}$, the cluster mass $M$, and the half-mass radius $r_{\text{h}}$. 
The \textsc{\scriptsize LIMEPY} code generated several profiles, such as the surface mass density $\Sigma(r_i)$, line-of-sight mean-square velocity $u^2_\textsc{\scriptsize L}(r_i)$, radial and tangential mean-square velocity on the projected plane $u^2_\textsc{\scriptsize R}(r_i)$ and $u^2_\textsc{\scriptsize T}(r_i)$.
Thus, the value of $\bar{\sigma}_\text{los}(r_i)$ is simply the square root of $u_\textsc{\scriptsize L}^2(r_i)$,
and $\bar{\sigma}_\text{pm}(r_i)$ is the square root of  $[u^2_\textsc{\scriptsize R}(r_i)+u^2_\textsc{\scriptsize T}(r_i)]/2$.

To complete the data-model fitting, two more parameters were needed. 
The cluster distance $D$ is a parameter that converts the radial coordinate of the theoretical profile from pc to arcsec and the observational proper-motion velocity dispersion from mas/yr to km/s.
The V-band mass-to-light ratio $\Upsilon$ is a parameter for producing the luminosity density $\Sigma(r_i)/\Upsilon$, 
and the surface brightness $\bar{\mu}(r_i)$ can be obtained by 
\begin{equation}
\bar{\mu}(r_i) = M_{\textsc{\scriptsize V},\odot}-5(1+\log c)-2.5\log(\Sigma(r_i)/\Upsilon),
\end{equation}
where $M_{\textsc{\scriptsize V}, \odot}=4.83$ mag is the V-band absolute magnitude of the Sun and $c=\pi/648000$ rad/arcsec is a factor for the unit conversion \citep{Watkins2015b}. 

Through the minimization of the $\chi^2$ function, the best-fit values of seven parameters 
$W_0$, $g$, $\hat{r}_\text{a}$, $M$, $r_{\text{h}}$, $D$, $\Upsilon$ can be obtained. 
We used the code \textsc{\scriptsize EMCEE} \citep{Foreman-Mackey2013} to perform the $\chi^2$ minimization.
It is an affine-invariant ensemble sampler that employs the Markov chain Monte Carlo (MCMC) process \citep{Goodman2010}.
One has to decide the initial distribution and the parameters range for the \textsc{\scriptsize EMCEE} samples.
For the concentration parameter $W_0$, the range was set to $1<W_0<15$. 
It covers a similar range in Table II of \citet{King1966} 
and represents various degrees of concentration of star clusters.
Figure 4 in \citet{Gieles2015} showed the relevant models for star clusters and the corresponding parameters; hence we set $0<g<3$ for the truncation parameter accordingly.
The dimensionless anisotropy radius $\hat{r}_{\text{a}}$ needs a wide range to include the isotropic models.
Therefore, we set a large range for $\log{\hat{r}_{\text{a}}}$ as $-1 < \log{\hat{r}_{\text{a}}} <20$.
For the remained parameters, we checked the literature values and considered wider ranges to include more possibilities. 
The ranges of these parameters were set to be $0.1<M<50$ ($10^5 \: \mbox{M}_{\odot}$), $0.1<r_{\text{h}}<15$ (pc), 
$0.1<D<35$ (kpc), and $0.1< \Upsilon <5$ ($\Upsilon_{\odot}$).
Finally, the initial distributions of all parameters are set to be uniform.

\section{Results and Discussion}


The best-fit results are displayed in Table~\ref{tab:theta}.
The first column shows the names of the clusters.
Seven fitting parameters are listed from the second to eighth columns.
The second column presents the concentration parameter $W_0$ and the values range roughly from 3 to 9 for these clusters.
The third and the fourth columns show the truncation parameter $g$ and the logarithm of the dimensionless anisotropy radius $\log{\hat{r}_{\text{a}}}$.
The fifth and sixth columns list the cluster mass $M$ and the half-mass radius $r_{\text{h}}$.
These clusters have $r_{\text{h}} \lesssim$ 10 pc. 
Among them, NGC 5139 has the largest mass and radius.
The heliocentric distance $D$ is shown in the seventh column.
Most clusters have $D \lesssim$ 12 kpc except for NGC 6715, which is roughly two times distant.
The eighth column reveals the V-band mass-to-light ratio $\Upsilon$.
To understand the anisotropy conveniently, the quantity $\kappa$ is shown in the ninth column.
NGC 5139 and NGC 7078 have $\kappa>1$, indicating the anisotropic behavior.
The quantity in the last column is the reduced chi-square $\chi^2_{\text{r}}$ defined by
\begin{equation}
\chi^2_{\text{r}}=\frac{\chi^2}{n-n_{\text{p}}},
\end{equation}
where $n$ is the total number of data and $n_{\text{p}}$ is the number of parameters.

\begin{table*}
\small
\caption{The properties of the clusters. The first column lists the names of the clusters. Columns two to eight show the fitting parameters, which are concentration parameter $W_0$, truncation parameter $g$, the logarithm of the dimensionless anisotropy radius $\log{\hat{r}_{\text{a}}}$, cluster mass $M$, half-mass radius $r_{\text{h}}$, distance $D$, and V-band mass-to-light ratio $\Upsilon$.
Column nine presents the quantity $\kappa$ which measures the amount of anisotropy, and the final column gives $\chi^2_{\text{r}}$.}
{\renewcommand{\arraystretch}{1.2}
\label{tab:theta}
\begin{tabular}{lccccccccc}
\\
\hline
cluster& $W_0$& $g$& $\log{\hat{r}_{\text{a}}}$& $M$&               $r_{\text{h}}$&          $D$&                   $\Upsilon$&      $\kappa$&                $\chi^2_{\text{r}}$ \\
       &      &    &                  & $(10^5 \: \mbox{M}_{\odot})$& $\mbox{(pc)}$&  $\mbox{(kpc)}$& $(\Upsilon_{\odot})$&                \\
\hline
NGC 104& $8.36\pm0.06$& $1.31\pm0.03$& $11.13_{-6.10}^{+6.02}$& $6.87\pm0.15$& $5.21\pm0.12$& $4.33\pm0.03$& $1.53\pm0.03$& 1.00& 2.10  \\
NGC 288& $4.46_{-0.82}^{+0.47}$& $1.55_{-0.38}^{+0.52}$& $10.40_{-6.41}^{+6.44}$& $1.02_{-0.10}^{+0.11}$& $8.26_{-0.32}^{+0.33}$& $9.80_{-0.36}^{+0.37}$& $2.32\pm0.12$& 1.00& 1.18 \\
NGC 362& $7.20\pm0.10$& $1.67\pm0.06$& $11.24_{-6.48}^{+5.87}$& $2.09_{-0.10}^{+0.11}$& $2.36_{-0.07}^{+0.08}$& $8.71_{-0.15}^{+0.16}$& $1.22\pm0.03$& 1.00& 4.74 \\
NGC 1851& $7.33_{-0.20}^{+0.19}$& $2.04\pm0.09$& $10.77_{-6.12}^{+6.26}$& $2.28_{-0.09}^{+0.10}$& $2.15_{-0.13}^{+0.15}$& $10.82_{-0.14}^{+0.15}$& $1.73_{-0.08}^{+0.09}$& 1.00& 1.61 \\
NGC 2808& $6.27_{-0.16}^{+0.17}$& $2.02_{-0.07}^{+0.10}$& $11.81_{-6.78}^{+5.01}$& $6.57_{-0.19}^{+0.25}$& $2.69_{-0.06}^{+0.08}$& $9.63_{-0.10}^{+0.12}$& $1.56_{-0.04}^{+0.05}$& 1.00& 1.63 \\
NGC 3201& $5.89_{-0.34}^{+0.31}$& $2.45\pm0.09$& $11.00_{-6.29}^{+6.19}$& $1.21_{-0.07}^{+0.08}$& $5.21_{-0.33}^{+0.42}$& $4.38_{-0.09}^{+0.10}$& $2.33_{-0.11}^{+0.12}$& 1.00& 2.74\\
NGC 5139& $4.02_{-0.65}^{+0.48}$& $1.94_{-0.26}^{+0.27}$& $0.41_{-0.10}^{+0.08}$& $32.82_{-0.67}^{+0.65}$& $8.82_{-0.17}^{+0.19}$& $5.32\pm0.03$& $2.38\pm0.09$& 1.15& 3.86 \\
NGC 5904& $7.03_{-0.10}^{+0.09}$& $1.56_{-0.04}^{+0.05}$& $10.39_{-6.07}^{+6.55}$& $3.03_{-0.15}^{+0.16}$& $4.54_{-0.11}^{+0.12}$& $7.24\pm0.13$& $1.39_{-0.03}^{+0.04}$& 1.00& 1.85 \\
NGC 6121& $7.52_{-0.13}^{+0.16}$& $0.46_{-0.21}^{+0.31}$& $9.80_{-5.59}^{+6.94}$& $0.81_{-0.04}^{+0.05}$& $3.20_{-0.13}^{+0.17}$& $1.85_{-0.03}^{+0.04}$& $2.11_{-0.08}^{+0.10}$& 1.00& 1.12 \\
NGC 6218& $5.77_{-0.35}^{+0.29}$& $1.51_{-0.22}^{+0.25}$& $10.69_{-6.53}^{+6.36}$& $0.75\pm0.06$& $3.02_{-0.13}^{+0.14}$& $4.59_{-0.14}^{+0.15}$& $1.78_{-0.08}^{+0.09}$& 1.00& 1.19 \\
NGC 6266& $7.84_{-0.09}^{+0.08}$& $0.62_{-0.10}^{+0.11}$& $10.90\pm6.23$& $5.98_{-0.24}^{+0.25}$& $2.55\pm0.08$& $6.33_{-0.08}^{+0.09}$& $1.85\pm0.05$& 1.00& 1.57 \\
NGC 6388& $7.09_{-0.11}^{+0.10}$& $1.68_{-0.08}^{+0.09}$& $10.86_{-6.13}^{+6.16}$& $7.79\pm0.20$& $2.07\pm0.05$& $10.35\pm0.10$& $1.68\pm0.03$& 1.00& 2.94 \\
NGC 6397& $9.17\pm0.17$& $0.87\pm0.08$& $10.96_{-6.11}^{+6.13}$& $0.79_{-0.03}^{+0.04}$& $3.73\pm0.19$& $2.40\pm0.04$& $2.47\pm0.12$& 1.00& 1.97\\
NGC 6441& $7.75\pm0.06$& $1.24_{-0.09}^{+0.10}$& $10.74_{-6.27}^{+6.31}$& $10.54_{-0.27}^{+0.28}$& $2.90_{-0.06}^{+0.07}$& $11.91\pm0.11$& $1.82\pm0.03$& 1.00& 3.35 \\
NGC 6656& $6.48_{-0.26}^{+0.23}$& $1.87_{-0.39}^{+0.34}$& $10.73_{-6.50}^{+6.33}$& $3.57_{-0.20}^{+0.22}$& $4.40_{-0.20}^{+0.29}$& $3.10\pm0.05$& $1.85\pm0.07$& 1.00& 1.04\\
NGC 6715& $6.99\pm0.07$& $2.21_{-0.03}^{+0.02}$& $11.26_{-6.28}^{+6.00}$& $17.79_{-1.06}^{+1.12}$& $5.28_{-0.25}^{+0.29}$& $25.08\pm0.53$& $2.07\pm0.06$& 1.00& 2.83\\
NGC 6752& $8.35_{-0.13}^{+0.12}$& $1.38\pm0.06$& $10.95_{-6.21}^{+6.16}$& $1.92\pm0.09$& $3.45\pm0.16$& $4.13\pm0.06$& $2.24\pm0.08$& 1.00& 1.20\\
NGC 7078& $8.30_{-0.13}^{+0.12}$& $0.86_{-0.13}^{+0.15}$& $1.16_{-0.06}^{+0.07}$& $5.08\pm0.17$& $4.05_{-0.17}^{+0.18}$& $10.40\pm0.12$& $1.53\pm0.05$& 1.16& 1.46\\
\hline
\end{tabular}
}
\end{table*}

\subsection{Comparison with Previous Work}

\begin{table*}
\small
\caption{The literature parameters of the clusters. The number in the parentheses represents the literature, (1) stands for \citet{Baumgardt2018}, (2) refers to \citet{Watkins2015b}, (3) corresponds to \citet{Baumgardt2021}, (4) represents \citet{Harris1996}, and (5) is \citet{Baumgardt2020}. The updated values for (1) and (5) are picked from the web catalog of \citet{Baumgardt2018}.}
{\renewcommand{\arraystretch}{1.2}
\label{tab:comp-theta}
\begin{tabular}{lcccccccc}
\\
\hline
cluster& $M$&                            $M$&                             $r_{\text{h}}$& $D$&            $D$&  $D$&  $\Upsilon$&  $\Upsilon$ \\
       & $(10^5 \: \mbox{M}_{\odot})$&    $(10^5 \: \mbox{M}_{\odot})$&     $\mbox{(pc)}$&  $\mbox{(kpc)}$& $\mbox{(kpc)}$&  $\mbox{(kpc)}$& $(\Upsilon_{\odot})$&   $(\Upsilon_{\odot})$  \\
 &  (1)&  (2)&  (1)&  (2)&  (3)&  (4)&  (5)&  (2) \\
\hline
NGC 104&   $8.95\pm0.06$&  $5.57_{-0.28}^{+0.33}$&  $6.30$&    $4.15\pm0.08$&  $4.521\pm0.031$&  $4.5$&  $1.96\pm0.09$&  $1.40\pm0.03$ \\ 
NGC 288&   $0.934\pm0.026$&  $0.79_{-0.11}^{+0.13}$&  $8.37$&    $9.03_{-0.56}^{+0.48}$&  $8.988_{-0.088}^{+0.089}$&  $8.9$&  $2.16\pm0.10$&  $2.20_{-0.10}^{+0.13}$  \\
NGC 362&   $2.84\pm0.04$&  ...&  $3.79$&  $...$&  $8.829\pm0.096$&  $8.6$&  $1.44\pm0.05$&  $...$  \\
NGC 1851&  $3.18\pm0.04$&  $1.78_{-0.11}^{+0.10}$&  $2.90$&  $10.32_{-0.24}^{+0.20}$&  $11.951_{-0.133}^{+0.134}$&  $12.1$&   $1.66\pm0.06$&  $1.51\pm0.03$  \\
NGC 2808&  $8.64\pm0.06$&  $5.91_{-0.25}^{+0.22}$&  $3.89$&  $9.45_{-0.15}^{+0.13}$&  $10.060_{-0.111}^{+0.112}$&  $9.6$&   $1.51\pm0.06$&  $1.56\pm0.02$  \\
NGC 3201&  $1.60\pm0.03$&  ...&  $6.78$&  $...$&  $4.737_{-0.042}^{+0.043}$&  $4.9$&   $2.16\pm0.09$&  $...$  \\
NGC 5139&  $36.4\pm0.4$&   $34.52_{-1.43}^{+1.45}$&  $10.36$&  $5.19_{-0.08}^{+0.07}$&  $5.426\pm0.047$&  $5.2$&   $2.58\pm0.10$&  $2.66\pm0.04$\\
NGC 5904&  $3.94\pm0.06$&  $3.65\pm0.75$&  $5.68$&  $7.79_{-0.61}^{+0.47}$&  $7.479\pm0.060$&  $7.5$&   $1.81\pm0.06$&  $1.43_{-0.10}^{+0.09}$  \\
NGC 6121&  $0.871\pm0.011$& ...&  $3.69$&  $...$&  $1.851_{-0.016}^{+0.015}$&  $2.2$&   $1.59\pm0.06$&  $...$  \\
NGC 6218&  $1.07\pm0.03$&  ...&  $4.05$&  $...$&  $5.109_{-0.048}^{+0.049}$&  $4.8$&   $1.92\pm0.09$&  $...$  \\
NGC 6266&  $6.10\pm0.04$&  $6.09_{-0.33}^{+0.39}$&  $2.43$&  $6.42\pm0.14$&  $6.412_{-0.104}^{+0.105}$&  $6.8$&   $1.99\pm0.11$&  $2.22\pm0.04$  \\
NGC 6388&  $12.5\pm0.1$&   $8.27_{-0.95}^{+0.89}$&  $4.34$&  $10.90_{-0.45}^{+0.40}$&  $11.171_{-0.161}^{+0.162}$&  $9.9$&   $2.19\pm0.06$&  $1.68_{-0.07}^{+0.06}$  \\
NGC 6397&  $0.966\pm0.013$&  $0.70_{-0.08}^{+0.09}$&  $3.90$&  $2.39_{-0.11}^{+0.13}$&  $2.482\pm0.019$&  $2.3$&   $1.66\pm0.07$&  $2.23_{-0.09}^{+0.10}$  \\
NGC 6441&  $13.2\pm0.1$&   ...&  $3.47$&  $...$&  $12.728_{-0.162}^{+0.163}$&  $11.6$&   $1.77\pm0.13$&  $...$  \\
NGC 6656&  $4.76\pm0.05$&  $2.49_{-0.37}^{+0.44}$&  $5.29$&  $2.84\pm0.16$&  $3.303\pm0.037$&  $3.2$&   $2.05\pm0.08$&  $1.88_{-0.10}^{+0.12}$  \\
NGC 6715&  $17.8\pm0.3$&   $11.83_{-0.53}^{+0.62}$&  $5.20$&  $22.57_{-0.39}^{+0.44}$&  $26.283_{-0.325}^{+0.328}$&  $26.5$&  $2.10\pm0.12$&  $1.94\pm0.03$  \\
NGC 6752&  $2.76\pm0.04$&  $1.82\pm0.12$&  $5.27$&  $4.02_{-0.08}^{+0.10}$&  $4.125\pm0.041$&  $4.0$&   $2.34\pm0.11$&  $2.14_{-0.06}^{+0.05}$  \\
NGC 7078&  $6.33\pm0.07$&  $4.95\pm0.19$&  $4.30$&  $10.36_{-0.16}^{+0.15}$&  $10.709_{-0.095}^{+0.096}$&  $10.4$&   $1.58\pm0.10$&  $1.49\pm0.02$  \\
\hline
\end{tabular}
}
\end{table*}

To compare our results with the previous work,
we used the measurable physical properties estimated in the published literature, as listed in Table~\ref{tab:comp-theta}.
We first considered the comparison of the cluster's total mass. 
In general, the masses estimated by \citet{Baumgardt2018} are larger than those estimated by \citet{Watkins2015b}, and our results are usually between their values.
Almost all of our results are very close to the masses estimated in \citet{Watkins2015b}.

We also compared our half-mass radius with the one in the catalog of \citet{Baumgardt2018}.
Generally, our results are smaller, consistent with the results of total mass, since our masses are lower than those in \citet{Baumgardt2018}.
Therefore, the radii of the clusters tend to be smaller to fit the line-of-sight velocity dispersion.
Some differences between the radius might come from the mass spectrum.
The radial distributions of different species may introduce additional variation between the half-mass radii.
Nevertheless, the mass-to-light ratios obtained in our work are consistent with the values in \citet{Baumgardt2020} and \citet{Watkins2015b}.

For distance comparison, we compared with the values in \citet{Watkins2015b}, \citet{Baumgardt2021}, and \citet{Harris1996}.
\citet{Watkins2015b} derived the distance by comparing their proper motion velocity dispersion with the line-of-sight velocity dispersion from the literature.
\citet{Baumgardt2021} calculated the mean distance from several methods, such as the Gaia EDR3 parallaxes, the method by fitting nearby subdwarfs to globular cluster main sequences, the color-magnitude diagram fitting, and the distances from the period-luminosity relation of RR Lyrae stars.
The distances in \citet{Harris1996} are a compilation of the distance measurements from the literature.

\begin{figure*}
    \centering
	\includegraphics[width=2\columnwidth]{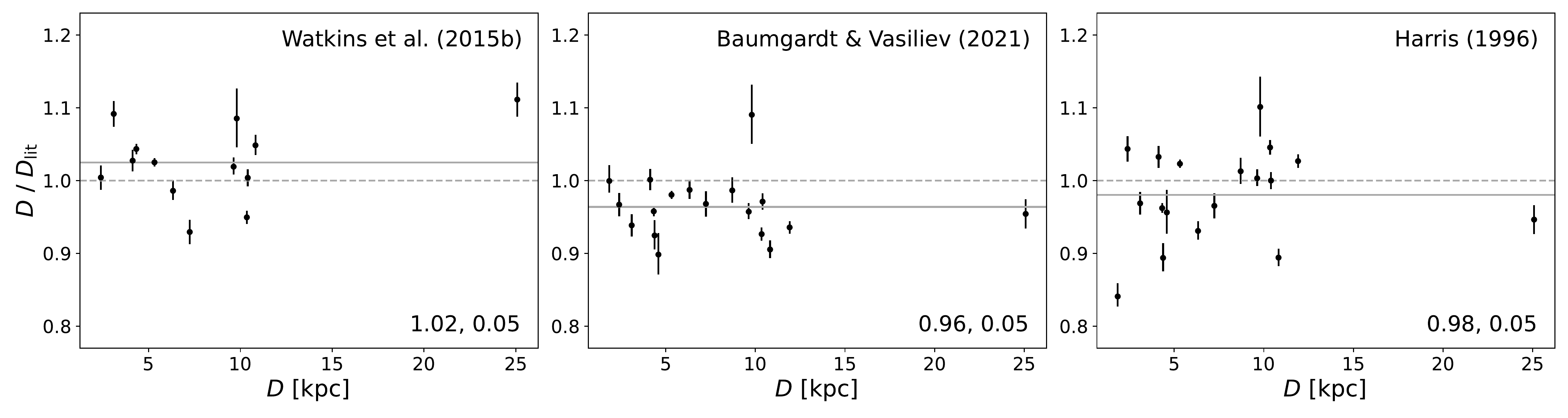}
    \caption{The comparison of the cluster distance with the values mentioned in earlier studies. The horizontal axis is the cluster distance obtained in this work, and the vertical axis shows the ratio of our value to the distance given in the literature. The dashed line and the solid line represent the unity and the average. Each panel is for comparison with the particular publication, as 
    labeled at the top-right corner.
    At the bottom-right corner, the left number is the averaged $D/D_{\text{lit}}$, and the right number is 
    the averaged $|D/D_{\text{lit}}-1|$.}
    \label{fig:comp-D}
\end{figure*}

Fig.~\ref{fig:comp-D} shows the ratio of our distance $D$ and the one published in literature $D_{\text{lit}}$, i.e., $D/D_{\text{lit}}$, for each considered cluster.
For each panel, the compared literature is labeled at the top-right corner.
Each point represents a particular cluster studied in the compared literature and this work.
The dashed line represents the unity, and the solid line is the average value of the ratio.
Two numbers are shown in the bottom-right of the panels, the left number is the averaged $D/D_{\text{lit}}$, and the right one is the averaged $|D/D_{\text{lit}}-1|$.
These numbers indicate that our results are closer to \citet{Harris1996} and \citet{Watkins2015b}, and slightly lower than \citet{Baumgardt2021}.
In general, our results agree with the values from these studies.

\subsection{The Profiles}

\begin{figure*}
    \centering
	\includegraphics[width=2\columnwidth]{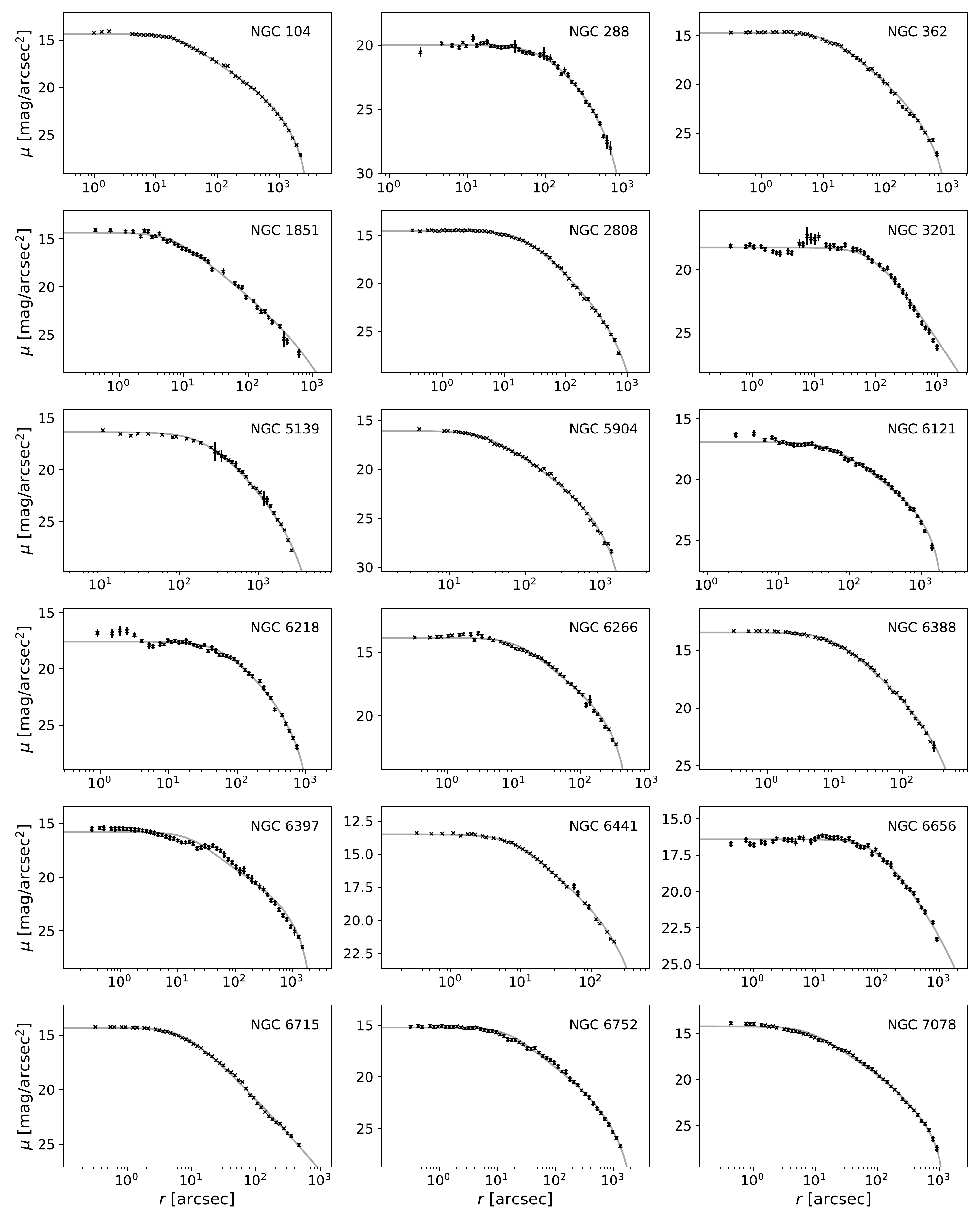}
    \caption{
    The surface brightness profiles of the clusters.
    The observations are shown as crosses, and the models are expressed by grey lines.
    For each panel, the name of the cluster is mentioned at the top-right corner. 
    }
    \label{fig:sb-r}
\end{figure*}

\begin{figure*}
    \centering
	\includegraphics[width=2\columnwidth]{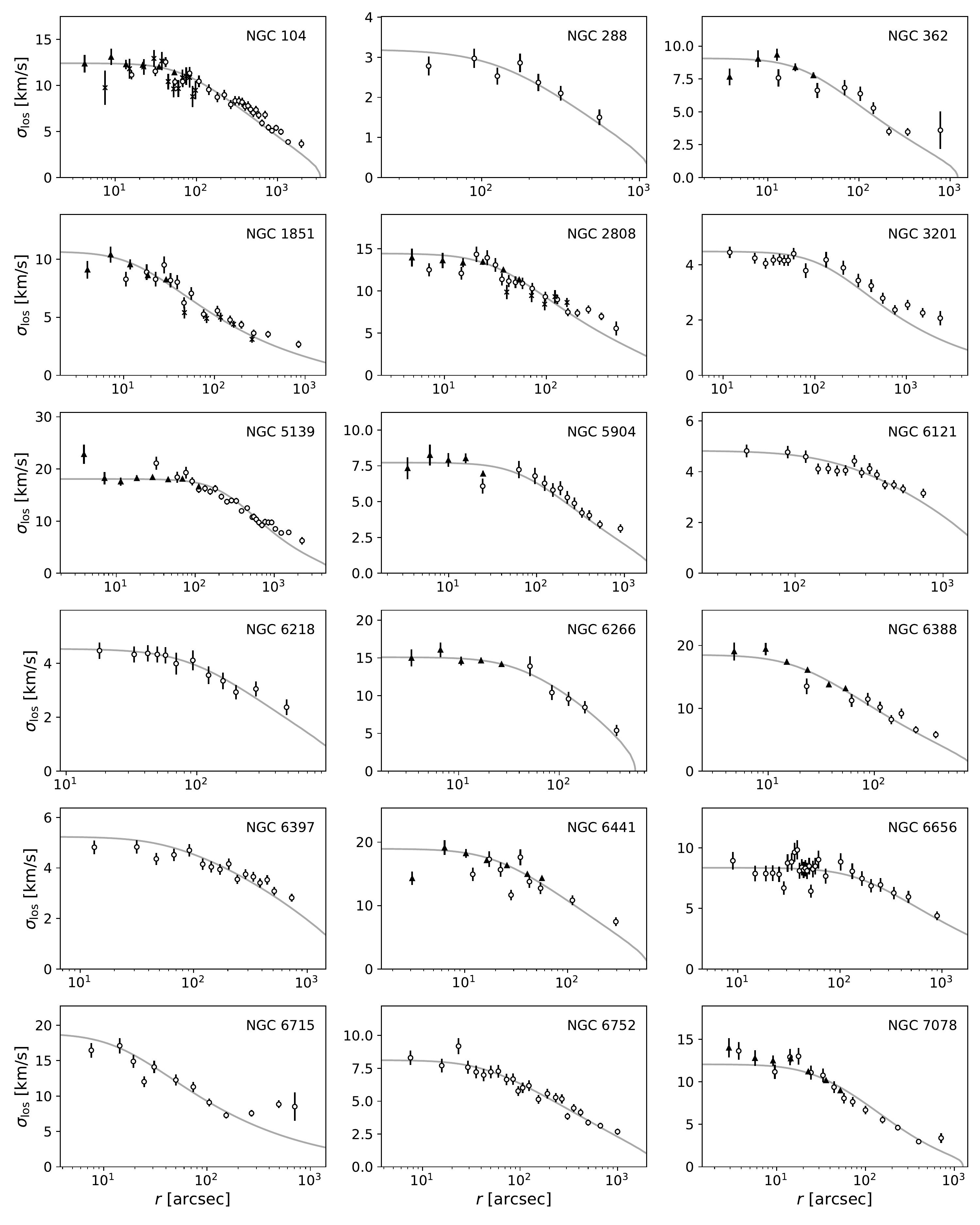}
    \caption{
    The line-of-sight velocity dispersion profiles of the clusters.
    The open circles represent the data of \citet{Baumgardt2017}, \citet{Baumgardt2018}, and \citet{Dalgleish2020}.
    The data of \citet{Kamann2018} are shown in solid triangles.
    The crosses are used for additional data of some clusters mentioned in Section 3.
    The models are expressed by grey lines.
    For each panel, the name of the cluster is mentioned at the top-right corner. 
    }
    \label{fig:Vlm-r}
\end{figure*}

\begin{figure*}
    \centering
	\includegraphics[width=2\columnwidth]{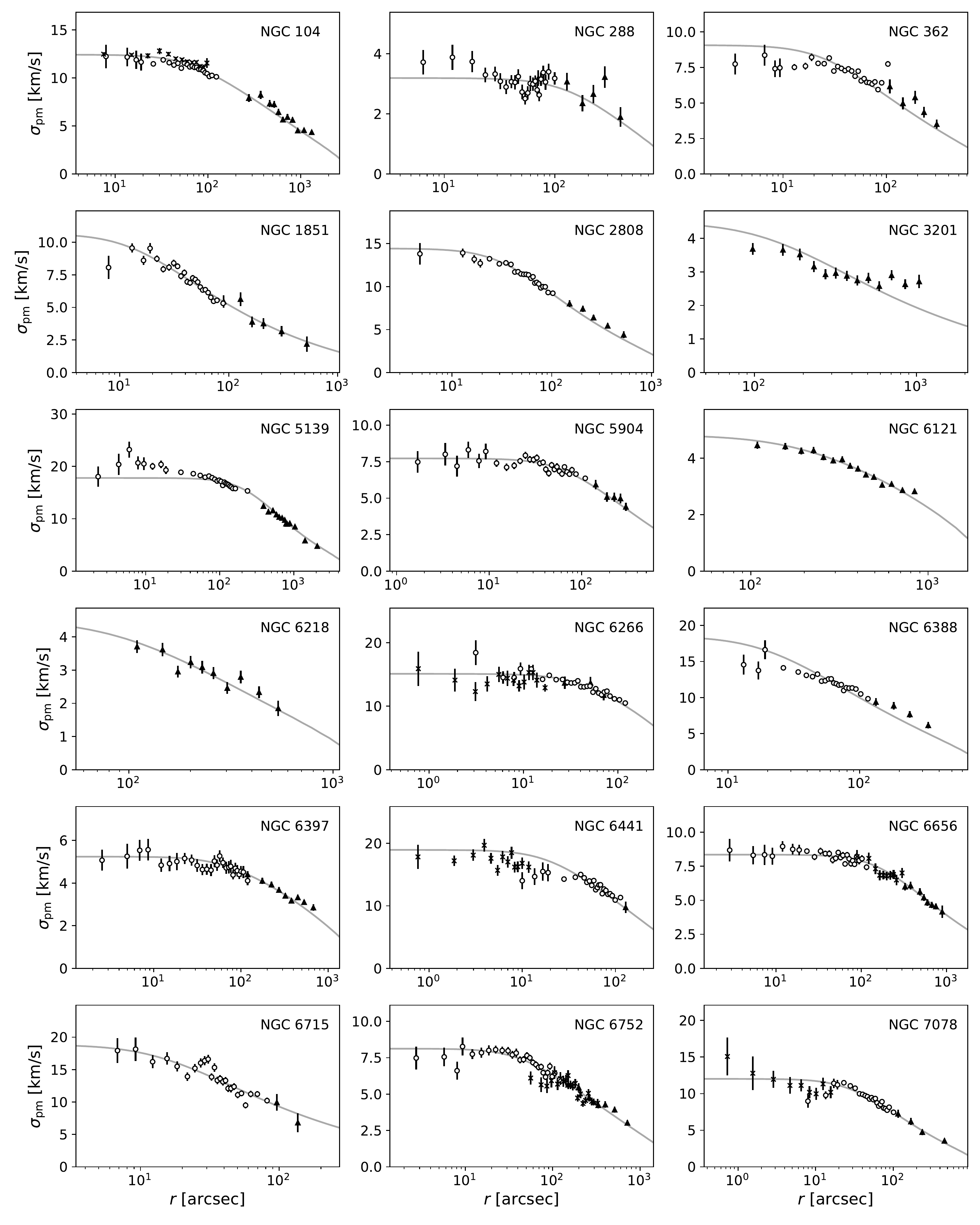}
    \caption{
    The proper-motion velocity dispersion profiles of the clusters.
    The open circles represent the data of \citet{Watkins2015a}.
    The data of \citet{Vasiliev2021} is shown in solid triangles.
    The crosses are used for additional data of some clusters mentioned in Section 3. 
    The models are expressed by grey lines.
    For each panel, the name of the cluster is mentioned at the top-right corner. 
    }
    \label{fig:Vpm-r}
\end{figure*}

Fig.~\ref{fig:sb-r} to~\ref{fig:Vpm-r} show the profiles of surface brightness, line-of-sight velocity dispersion, 
and proper-motion velocity dispersion.
The horizontal axis is the distance from the cluster's center in arcsec.
The vertical axis gives the surface brightness in mag/arcsec$^2$ in Fig.~\ref{fig:sb-r}, and velocity dispersion in km/s from Fig.~\ref{fig:Vlm-r} to~\ref{fig:Vpm-r}.
It can be seen that \textsc{\scriptsize LIMEPY} models can produce similar profiles as observational ones.
To examine these clusters more quantitatively, we classified the results by $\chi^2_{\text{r}}$.
Many clusters were found to have $\chi^2_{\text{r}}<2$. 
These clusters have suitable fittings for all three profiles, as shown in the figures.

NGC 362 has the largest $\chi^2_{\text{r}}$, and the model profiles agree with the observations in surface brightness and line-of-sight velocity dispersion.
However, the central part of the modeled proper-motion velocity dispersion is slightly larger than the observations.
Data with a small error bar in the outer part located much higher than the profile, making the fitting worse.
NGC 6441 also has a larger $\chi^2_{\text{r}}$.
The model agrees well with the surface brightness and the outer part of the proper motion velocity dispersion but predicts larger values for the inner part.
The model can also fit the rough trend of the line-of-sight velocity dispersion, but some points lie below the model.

For NGC 3201, the model has smaller line-of-sight velocity dispersion for radius above 100 arcsec.
There are also some under estimations for the proper motions in the outermost region, 
where the observational profile tends to level off rather than continue to decrease.
Some scenarios were proposed to explain the higher velocity dispersion in the outer part, such as the orbital history with accretion and the embedding by a dark matter halo  \citep{Bianchini2019}. 
It was also found that binary stars could contribute to part of the effect \citep{Wan2021}.
For NGC 6715, the model agrees with the observations, except for the outermost region of the line-of-sight velocity dispersion, where the observational profile grows.
This rise is probably caused by the stars in the nucleus of the Sagittarius dwarf galaxy, where NGC 6715 inhabits \citep{Bellazzini2008}.

NGC 5139 has large central velocity dispersions, which the model cannot explain well.
For NGC 6388, the model has a steeper proper-motion velocity dispersion profile than the observational one.
Further discussions of these two clusters will be made in the following subsection.

\subsection{Possible Intermediate-Mass Black Hole ?}

Stellar black holes exist in astrophysical systems such as X-ray binaries \citep{Mikolajewska2022}. 
In addition, supermassive black holes are also confirmed to exist at the centers of our Milky Way \citep{GRAVITY2019} and other galaxies \citep{Blandford2019}.
Whether there are any intermediate-mass black holes in the universe is one of the most important questions in astronomy. 
Globular clusters are considered good candidates to host intermediate-mass black holes and thus attract much attention. 
Among 18 globular clusters in the present work, 
NGC 5139 was discussed previously as a likely candidate.

\begin{figure*}
    \centering
	\includegraphics[width=2.0\columnwidth]{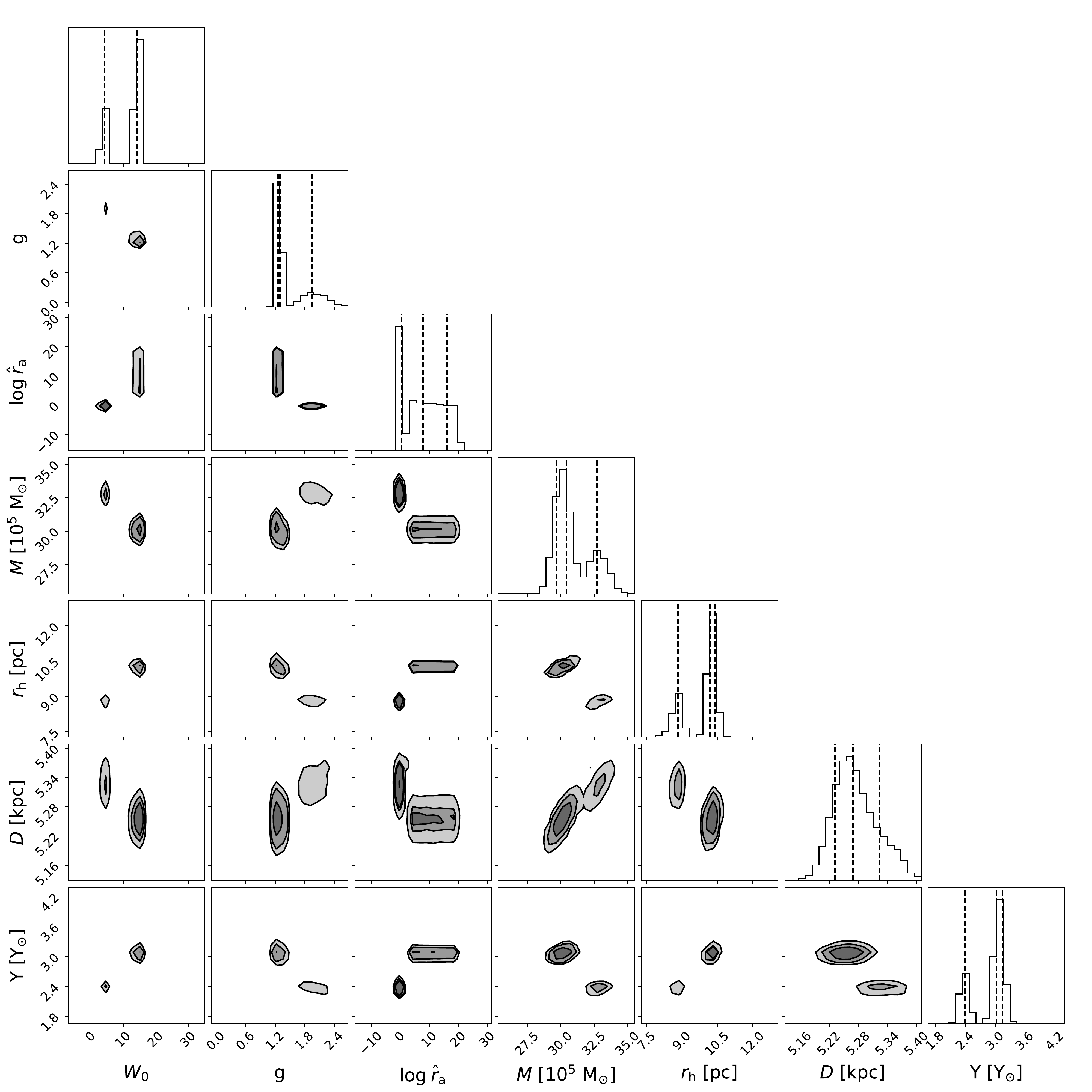}
    \caption{The MCMC posterior parameter distributions of NGC 5139.}
    \label{fig:5139-corner}
\end{figure*}

For our work here, the data-model fitting of NGC 5139 led to two groups of model parameters, as shown in Fig.~\ref{fig:5139-corner}.
These groups have very different concentration parameters $W_0$ and logarithm of the dimensionless anisotropy radius $\log{\hat{r}_{\text{a}}}$.
One has smaller $W_0$ and $\log{\hat{r}_{\text{a}}}$, and the other has larger values. 
Hence, we do further fittings with narrower ranges as $1<W_0<8$, $-1 < \log{\hat{r}_{\text{a}}} <2$,
and $8<W_0<15$, $2 < \log{\hat{r}_{\text{a}}} <20$, separately.
The results are shown in Table~\ref{tab:5139-models}.
We denote the one with lower $\chi^2_{\text{r}}$ as Model A,  the result previously listed in Table~\ref{tab:theta} 
and presented in Fig.~\ref{fig:sb-r} to~\ref{fig:Vpm-r}.
Model A has a low concentration. 
It also has a small dimensionless anisotropy radius with $\kappa = 1.15$, making it more anisotropic. 
In contrast, Model B is isotropic with a high concentration.

\begin{figure*}
    \centering
    \includegraphics[width=1.3\columnwidth]{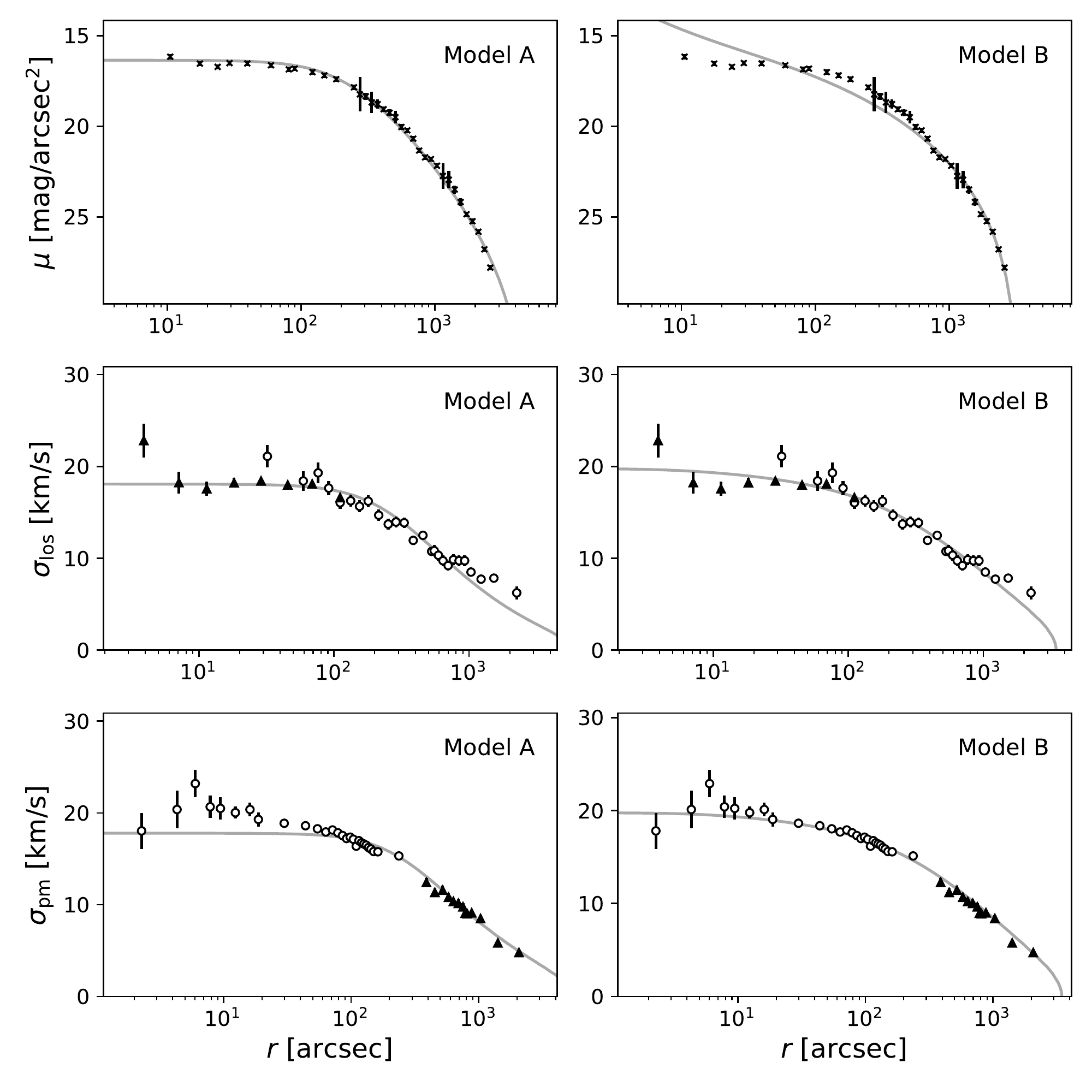}
    \caption{The comparison of the profiles from two models of NGC 5139. Left panels show the results from Model A and the right panels show those from Model B.
    The open circles represent the data of \citet{Baumgardt2017}, \citet{Baumgardt2018}, and \citet{Dalgleish2020} for line-of-sight velocity dispersions and \citet{Watkins2015a} for proper motion velocity dispersions. The solid triangles are used for the data of \citet{Kamann2018} for line-of-sight velocity dispersions and \citet{Vasiliev2021} for proper motion velocity dispersions.
    The models are expressed by grey lines.
    }
    \label{fig:5139-models}
\end{figure*}

The best-fit profiles are shown in Fig.~\ref{fig:5139-models}.
Model A fits the surface brightness well but predicts lower central velocity dispersion, especially for the proper motion.
On the other hand, Model B has good fittings on both velocity dispersion but a poor fitting on the surface brightness.
The deviation in surface brightness leads to a larger $\chi^2_{\text{r}}$.
Although the data and radial range of the observational kinematic profiles differs, 
the parameters from Model A agree with those in the best-fit model in \citet{Zocchi2017}.

These results obtained with two models show that it is difficult to perfectly and simultaneously fit all profiles of NGC 5139  with the current considered model. 
This could indicate the existence of central dark objects which can cause an increase in central velocities.
These objects could be an intermediate-mass black hole \citep{Noyola2010, Baumgardt2017} or a group of stellar-mass black holes at the cluster center \citep{Baumgardt2019b}. 
Both can also suppress the mass segregation of the stars \citep{Gill2008, Peuten2016} and render the cluster to have a larger core \citep{Baumgardt2005, Peuten2017}.
The main difference is that the intermediate-mass black hole could produce some stars faster than 60 km/s in the central 20 arcsec of NGC 5139, 
which was not confirmed in current observations \citep{Baumgardt2019b}.

NGC 6388 is another candidate cluster that may host a central intermediate-mass black hole.
The study of the integrated light spectra revealed a high central LOS velocity dispersion $\sim$25 km/s within 2 arcsec \citep{Lutzgendorf2011}.
However, there was also a result that suggests a dispersion $\sim$15 km/s in the same region derived from stars' radial velocities \citep{Lanzoni2013}.
Hence, the actual kinematic behavior of the cluster center is not clear.
The data we used have the extension to nearly 5 arcsec with a velocity dispersion $\sim$20 km/s.
Our results show that the surface brightness and line-of-sight velocity dispersion can be fitted well without the central black hole.
However, the model predicts a steeper proper-motion velocity-dispersion profile than the observations, being higher inside but lower outside.
This behavior can also be seen in Figure 9 of \citet{Watkins2015b}.

NGC 7078 is also a candidate cluster that could host an intermediate-mass black hole.
The increase in central velocity dispersion found in Hubble Space Telescope was explained by an intermediate-mass black hole \citep{Gerssen2002}.
However, the cluster can also be fitted with a group of dark stellar remnants \citep{denBrok2014} or N-body simulations without intermediate-mass black holes \citep{Baumgardt2017}.
In our results, the cluster could be fitted well without central black holes, and some degree of anisotropy was observed, which can raise the central velocities.
In addition, although there are raised velocity dispersions in observation, the uncertainties of the data are also large.
Therefore, we obtain a better fitting than NGC 5139.

\begin{table*}
\small
\caption{The parameters of two models of NGC 5139. The first column indicates different models. The second to eighth columns show the fitting parameters. The quantities in the last two columns are $\kappa$ and $\chi^2_{\text{r}}$.}
{\renewcommand{\arraystretch}{1.2}
\label{tab:5139-models}
\begin{tabular}{lccccccccc}
\\
\hline
Model& $W_0$& $g$& $\log{\hat{r}_{\text{a}}}$& $M$&               $r_{\text{h}}$&          $D$&                   $\Upsilon$&      $\kappa$&              $\chi^2_{\text{r}}$ \\
       &      &    &                  & $(10^5$ $\mbox{M}_{\odot})$& $\mbox{(pc)}$&  $\mbox{(kpc)}$& $(\Upsilon_{\odot})$&                \\
\hline
A& $4.02_{-0.65}^{+0.48}$& $1.94_{-0.26}^{+0.27}$& $0.41_{-0.10}^{+0.08}$& $32.82_{-0.67}^{+0.65}$& $8.82_{-0.17}^{+0.19}$& $5.32\pm0.03$& $2.38\pm0.09$& 1.15& 3.86 \\
B& $14.16_{-0.23}^{+0.25}$& $1.28\pm0.03$& $11.38_{-6.00}^{+5.83}$& $30.10\pm0.59$& $10.26_{0.16}^{0.17}$& $5.25\pm0.03$& $3.07\pm0.09$& 1.00& 5.64 \\
\hline
\end{tabular}
}
\end{table*}

\subsection{The Anisotropy}
Two clusters, NGC 5139 and NGC 7078, possess small dimensionless anisotropy radius and reveal some degree of anisotropy.
The former has $\kappa = 1.15$ and the latter has $\kappa = 1.16$.
Other clusters have isotropic behavior with $\kappa = 1.00$ and a large anisotropy radius.
One effect of radial anisotropy is that it can increase the central velocity dispersion.
The rise in central velocity dispersions can be seen in Fig.~\ref{fig:Vlm-r} and~\ref{fig:Vpm-r}.
On the other hand, the amount of anisotropy estimated from our fittings could be underestimated, since the difference between tangential and radial proper motions will be averaged out in the combined proper motion velocity dispersion.

The results are reasonable compared with some previous studies.
For example, the parameters of NGC 5139 are similar to those estimated in \citet{Zocchi2017} which the fittings were carried out with both radial and tangential proper motion velocity dispersions.
The weak anisotropy in many clusters were also reported by \citet{Watkins2015a} and \citet{Watkins2015b}, in which most of our samples were also studied.
\citet{Watkins2015b} showed that their distance estimation had good agreement with \citet{Harris1996} and concluded that the assumption of isotropy for their samples is reasonable.
\citet{Watkins2015a} examined the ratio $\sigma_\textsc{\scriptsize T}/\sigma_\textsc{\scriptsize R}$, which compared the tangential and radial components of the proper motion velocity dispersion at different radii.
They found that the cluster centers are relatively isotropic, and the behavior of the increasing anisotropy with the radius was very moderate.
From their figures, it can be seen that the decreasing of $\sigma_\textsc{\scriptsize T}/\sigma_\textsc{\scriptsize R}$ with a growing radius is more evident for NGC 5139 and NGC 7078.

In recent years, Gaia has provided the proper motion data in the outer parts of globular clusters, and the behavior of $\sigma_\textsc{\scriptsize T}/\sigma_\textsc{\scriptsize R}$ reveals more evidence of anisotropy \citep{Jindal2019, Vasiliev2021}.
In both studies, NGC 5904 appears to be isotropic, and NGC 104, NGC 5139, and NGC 7078 show radial anisotropy.
Some clusters are anisotropic in one study but are isotropic or uncertain in another; these include NGC 2808, NGC 6121, NGC 6397, NGC 6656, and NGC 6752.

\begin{figure*}
    \centering
	\includegraphics[width=2\columnwidth]{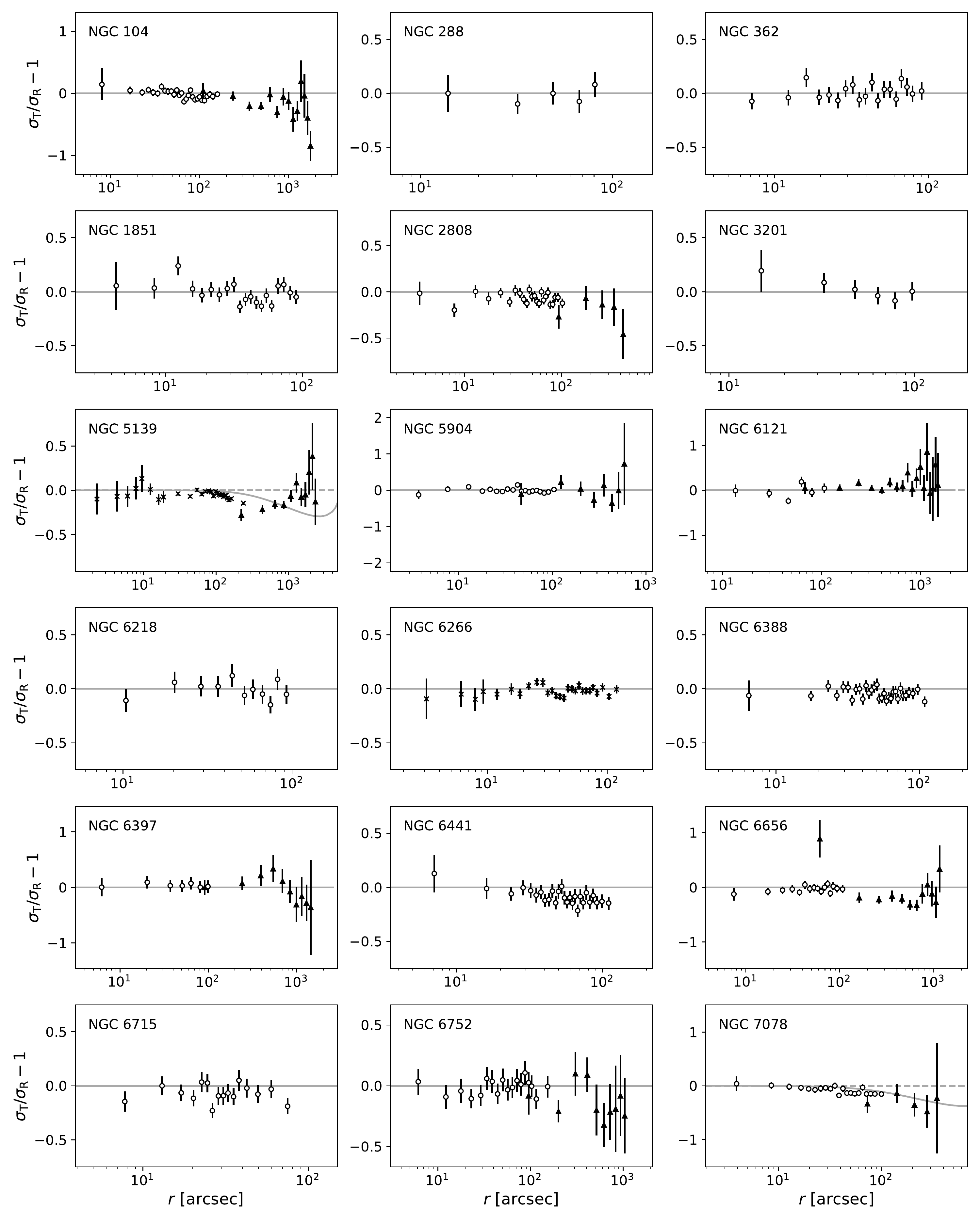}
    \caption{
    The anisotropy profiles of the clusters.
    The open circles represent the data of \citet{Libralato2022}.
    The data of \citet{Jindal2019} are shown in solid triangles.
    The crosses are used for \citet{Watkins2015a}.
    The models are expressed by grey lines.
    The horizontal grey dashed lines represent zeros which indicate isotropy.
    For each panel, the name of the cluster is mentioned at the top-left corner.}
    \label{fig:ani-r}
\end{figure*}

In addition, 
the anisotropy profiles $\sigma_\textsc{\scriptsize T}/\sigma_\textsc{\scriptsize R}-1$ from the observations and our models are plotted in Fig.~\ref{fig:ani-r}. 
The observational data was mainly from a recent report on the globular-cluster survey through Hubble Space Telescope \citep{Libralato2022}. 
It includes 16 clusters of our samples. 
The remaining two clusters were supplemented with the data from \citet{Watkins2015a}.
The data from Gaia \citep{Jindal2019} which contains half of our samples were also used.
In Fig.~\ref{fig:ani-r}, the data of the above-discussed literature are expressed by open circles, crosses, and solid triangles; the profiles are roughly isotropic or slightly radial anisotropic within $r\lesssim100$ arcsec.
The larger anisotropy appears mainly in the outer regions.
The radial anisotropy of NGC 5139 tends to increase from near 100 arcsec and later decrease to isotropy in $r\gtrsim1000$ arcsec.
Our model predicts the decrease in radial anisotropy at a larger radius.
For NGC 7078, the model shows a similar and milder profile to the observational one.
NGC 6121 shows isotropy inside but grows to tangential anisotropy at a larger radius.
The cluster was also found to be tangential anisotropy in \citet{Vasiliev2021}.
It could imply a more substantial influence from the tidal field, which is consistent with our results that this cluster has a smaller truncation parameter than others.

\subsection{The Imprint of Galactic Tidal Field}

As mentioned earlier, the truncation parameter has the effect of making the extent of the system finite, and also drives the profile to be isotropic near the edge.
These make the truncation parameter play a similar role as the external tidal field for the cluster.
The external field generally becomes weaker for a larger distance from the Galactic center.
Thus, clusters at larger distances from the Galactic center might be more extended and have larger values of truncation parameter $g$.

In addition, \citet{Chernoff1986} found that the tidal field can increase the evolution rate of the cluster through relaxation and shock heating.
Therefore, clusters closer to the Galactic center tend to evolve faster.
They also suggested that inner regions of the Galaxy could be good places to look for the core-collapsed clusters.
This agreed with \citet{Djorgovski1986} who found that the mean and median distances of core-collapsed clusters from the Galactic center are smaller than 5 kpc.

Moreover, the simulation in \citet{Zocchi2016} showed some related properties during the evolution of a globular cluster in an external tidal field.
For example, the truncation parameter $g$ and the cluster mass $M$ decrease during the evolution.
The concentration parameter $W_0$ grows with time and decreases slightly after core collapse.
The half-mass radius $r_{\text{h}}$ also increases with time and decreases as the cluster loses most of its mass.

\begin{figure}
    \centering
	\includegraphics[width=\columnwidth]{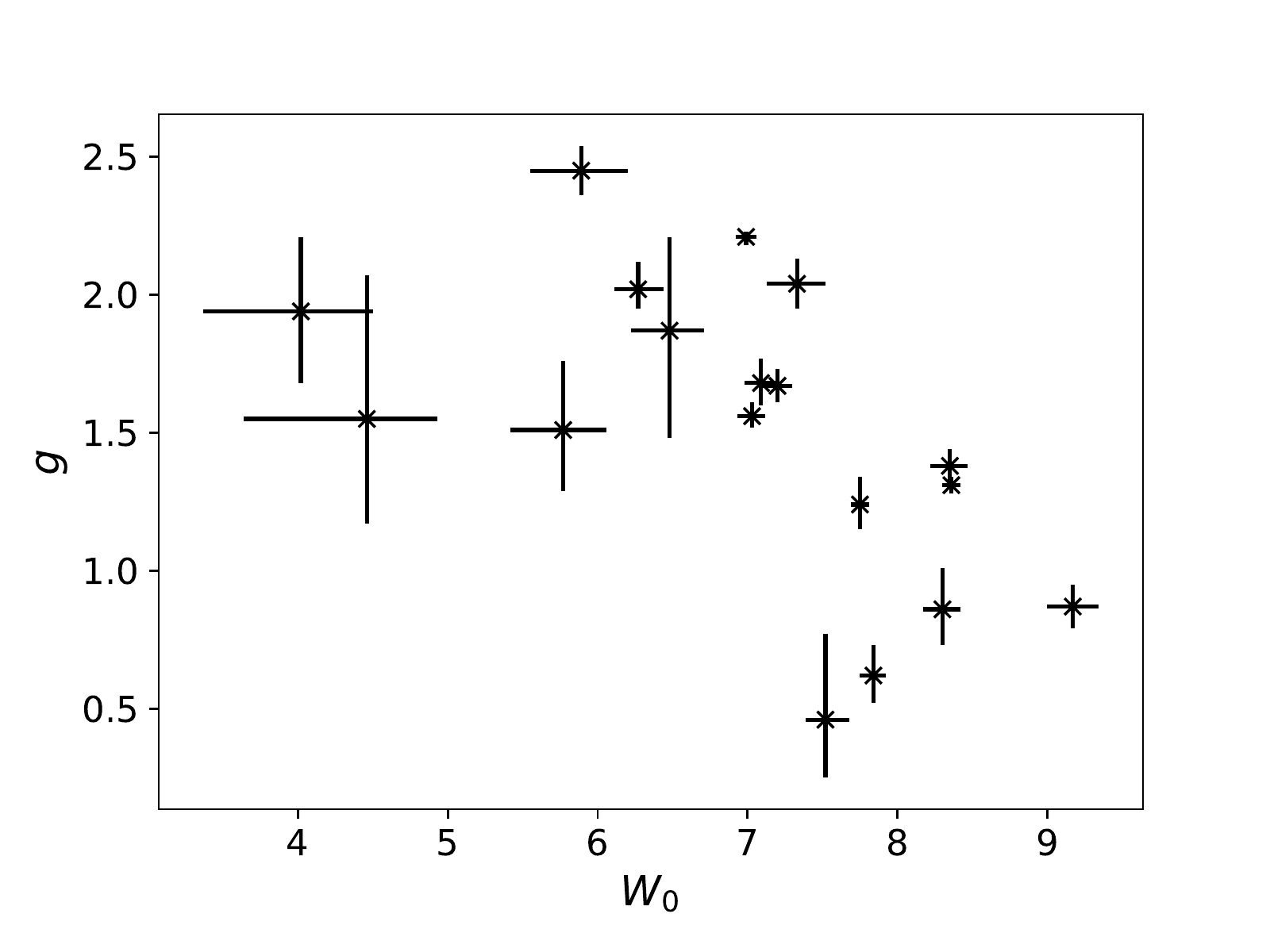}
    \caption{The truncation parameter versus the concentration parameter. The vertical axis represents the truncation parameter and the horizontal axis expresses the concentration parameter. Each point corresponds to a particular cluster.}
    \label{fig:w-g}
\end{figure}

\begin{figure}
    \centering
	\includegraphics[width=\columnwidth]{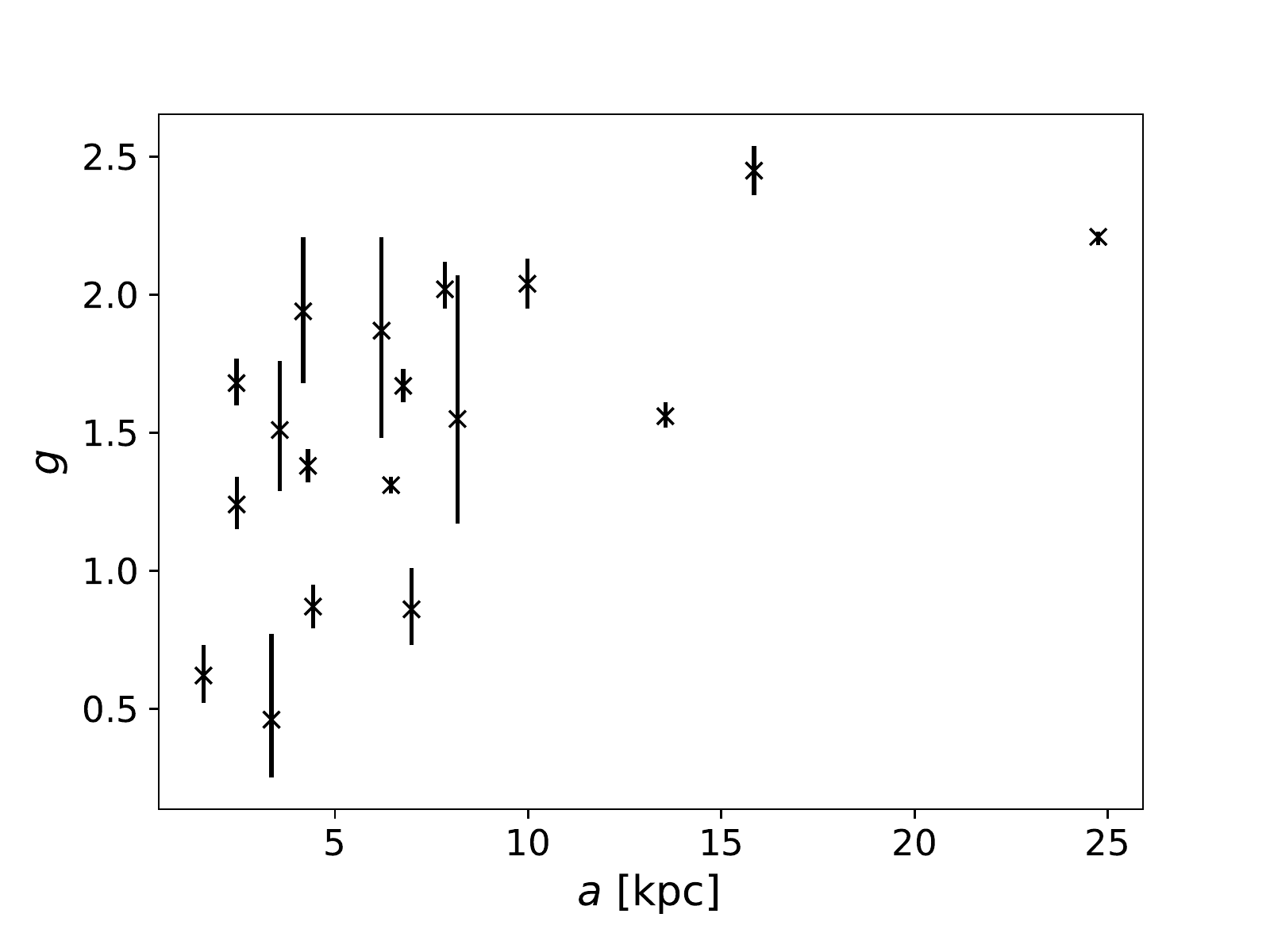}
    \caption{The truncation parameter versus the semimajor axis of the cluster. The vertical axis represents the truncation parameter and the horizontal axis expresses the semimajor axis. Each data point corresponds to a particular cluster.}
    \label{fig:a-g}
\end{figure}

Motivated by the above results, here we examine possible correlations between any pairs among the concentration parameter $W_0$, the truncation parameter $g$, the cluster mass $M$, the half-mass radius $r_{\text{h}}$, and the semimajor axis of the cluster orbit $a$. 
The values of $a$ were taken as the average of the apogalactic and perigalactic distances in \citet{Baumgardt2019a}, and the rest are our best-fit values in Table \ref{tab:theta}. 
The Spearman rank-order correlation coefficients, $C_{\rm s}$, were then calculated for all possible combinations; there were only two pairs with an absolute value of $C_{\rm s}$ greater than 0.5.
The first pair is the concentration parameter $W_0$ and the truncation parameter $g$. 
Their $C_{\rm s} = -0.65$ indicates a strong anti-correlation between $W_0$ and $g$. 
The distribution is presented in Fig.~\ref{fig:w-g}.
The second pair is the truncation parameter $g$ and the semimajor axis $a$ of the cluster orbit.
The corresponding correlation coefficient $C_{\rm s} = 0.60$ indicates a strong correlation between $g$ and $a$; the result is presented in Fig.~\ref{fig:a-g}.

The anti-correlation between the concentration parameter $W_0$ and the truncation parameter $g$ is reasonable, as 
those with smaller truncation parameters would have experienced stronger tidal fields and evolve faster. 
It is likely that a certain fraction of them become core-collapsed clusters and thus have larger concentrations. 
This anti-correlation is also consistent with the simulations in \citet{Zocchi2016}.
They showed that when the clusters form, 
the value of concentration parameter $W_0$ is nearly 4 and the value of truncation parameter $g$ is nearly 2.5. 
During the evolution, the truncation parameter $g$ decreases, but the concentration parameter $W_0$ increases. 
Therefore, in Fig.~\ref{fig:w-g},
younger clusters are located at the top-left corner, and the older clusters are distributed at the bottom-right corner.
However, the exact relationships between these two parameters for different clusters are still complicated and the
strength of this anti-correlation was not quantitatively investigated before. 

On the other hand, the correlation between the truncation parameter $g$ and the semimajor axis $a$ can be easily understood. 
The smaller truncation parameter shows that a stronger tidal field influences the cluster, and those clusters with smaller $a$ do experience stronger tidal fields. 
However, the relation between the two above-mentioned  parameters shall also depends on the initial size and the orbital evolution of a cluster.
The contribution from different Galactic components make the exact behavior of the tidal field more complicated.
It is reasonable that this correlation has a correlation coefficient $C_{\rm s} = 0.60$.

The strong $W_0 - g$ anti-correlation and $g - a$ correlation shall be regarded as 
observational results as the employed parameters are obtained
through our data-model fitting or from an observational catalog in literature. 
In addition, these observational anti-correlation and correlation agree with theoretical predictions.



\section{Summary and Conclusions}

In this work, we studied 18 clusters with the \textsc{\scriptsize LIMEPY} models, a unified family of isothermal models.
It can generate clusters with different amounts of concentration, truncation, and anisotropy, which are parametrized by continuous real numbers.
Including some current observational data, such as the MUSE survey and Gaia mission,
the fittings were carried out with a Markov Chain Monte Carlo ensemble sampler \textsc{\scriptsize EMCEE} 
and the parameters were determined by minimizing the $\chi^2$ of the fittings.

The measurable physical properties such as masses and distances, were compared with the values from the literature.
Usually, \citet{Baumgardt2018} has larger masses, while \citet{Watkins2015b} has smaller ones, and our results are in between.
The smaller half-mass radius in our results is consistent with the smaller mass estimated compared with \citet{Baumgardt2018}.
Some differences between the radius estimations might come from the effect of the mass spectrum.
For distance, our estimations are in agreement with the literature.
The mass-to-light ratios are also similar to the literature.

Generally, the models could produce profiles similar to the observational ones for most clusters.
For NGC 5139, there are two groups of parameters that correspond to a better fitting for the surface brightness or the velocity-dispersion profiles.
The anisotropic model gives a smaller $\chi^2_{\text{r}}$ and agrees with the best-fit results in \citet{Zocchi2017}. 
Some possible central dark objects, like an intermediate-mass black hole or a group of stellar-mass black holes might improve the fitting.
NGC 6388 is also a candidate to host an intermediate-mass black hole, with the actual central line-of-sight velocities being uncertain.
The data we used have the extension to nearly 5 arcsec with a velocity dispersion $\sim$20 km/s.
It could be fitted well with the \textsc{\scriptsize LIMEPY} model except for the slope of proper-motion velocity dispersion. 

For the anisotropy, NGC 5139 and NGC 7078 are anisotropic with $\kappa = 1.15$ and $\kappa = 1.16$.
The anisotropy leads to the rise in central velocity dispersion in these clusters.
Our estimations could have some underestimations because the data are combined proper motion dispersion profiles rather than separated radial and tangential profiles.
Nevertheless, the results are reasonable compared with some literature, such as \citet{Watkins2015a} and \citet{Watkins2015b}, where the anisotropy in the studied clusters seem small.

From a theoretical aspect, the truncation parameter may render the cluster to have a finite extension and isotropic profiles near the edge.
It is similar to the effect of the external tidal field.
In addition, a strong anti-correlation between the concentration parameter $W_0$ and the truncation parameter $g$ was confirmed,
which gives the imprint of the dynamical evolution of clusters.
Finally, a strong correlation between the truncation parameter $g$ and the semimajor axis $a$ was also found, which could result from the influence of the Galactic tidal field.

\section*{Acknowledgements}
We are grateful to the reviewer, Holger Baumgardt, for the useful suggestions which improved this paper significantly.
We acknowledge the financial support from the Ministry of Science and Technology, Taiwan, (MOST grant 110-2112-M-007-035).
We are grateful to the authors of Trager et al. (1995), Harris (1996), McLaughlin \& van der Marel (2005), Baumgardt (2017), Baumgardt \& Hilker (2018), Dalgleish et al. (2020), Kamann et al. (2018), McLaughlin et al. (2006), Watkins et al. (2015a), Vasiliev \& Baumgardt (2021), Häberle et al. (2021), McNamara et al. (2003), McNamara et al. (2012), Zloczewski et al. (2012), Watkins et al. (2015b), Baumgardt \& Vasiliev (2021), Baumgardt et al. (2020), Libralato et al. (2022), Jindal et al. (2019), Baumgardt et al. (2019a), for making their data publicly available. 
This paper used the VizieR catalogue access tool, operated at CDS, Strasbourg, France, and the Astrophysics Data System Bibliographic Services of National Aeronautics Space and Administration, USA.
{\it Software}: \textsc{\scriptsize LIMEPY} (Gieles \& Zocchi 2015), \textsc{\scriptsize EMCEE} (Foreman-Mackey et al. 2013), corner, NumPy, and SciPy.

\section*{Data Availability}
The electronic file of Table 1 is available in machine-readable form at VizieR (vizier.u-strasbg.fr) of Strasbourg astronomical Data Center (CDS).


\bibliographystyle{mnras}
\bibliography{refpaper} 

\begin{thebibliography}{}
\makeatletter
\relax
\def\mn@urlcharsother{\let\do\@makeother \do\$\do\&\do\#\do\^\do\_\do\%\do\~}
\def\mn@doi{\begingroup\mn@urlcharsother \@ifnextchar [ {\mn@doi@}
  {\mn@doi@[]}}
\def\mn@doi@[#1]#2{\def\@tempa{#1}\ifx\@tempa\@empty \href
  {http://dx.doi.org/#2} {doi:#2}\else \href {http://dx.doi.org/#2} {#1}\fi
  \endgroup}
\def\mn@eprint#1#2{\mn@eprint@#1:#2::\@nil}
\def\mn@eprint@arXiv#1{\href {http://arxiv.org/abs/#1} {{\tt arXiv:#1}}}
\def\mn@eprint@dblp#1{\href {http://dblp.uni-trier.de/rec/bibtex/#1.xml}
  {dblp:#1}}
\def\mn@eprint@#1:#2:#3:#4\@nil{\def\@tempa {#1}\def\@tempb {#2}\def\@tempc
  {#3}\ifx \@tempc \@empty \let \@tempc \@tempb \let \@tempb \@tempa \fi \ifx
  \@tempb \@empty \def\@tempb {arXiv}\fi \@ifundefined
  {mn@eprint@\@tempb}{\@tempb:\@tempc}{\expandafter \expandafter \csname
  mn@eprint@\@tempb\endcsname \expandafter{\@tempc}}}

\bibitem[\protect\citeauthoryear{{Bailyn}}{{Bailyn}}{1995}]{Bailyn1995}
{Bailyn} C.~D.,  1995, \araa, \href
  {https://ui.adsabs.harvard.edu/abs/1995ARA&A..33..133B} {33, 133}

\bibitem[\protect\citeauthoryear{{Baumgardt}}{{Baumgardt}}{2017}]{Baumgardt2017}
{Baumgardt} H.,  2017, \mn@doi [\mnras] {10.1093/mnras/stw2488}, \href
  {https://ui.adsabs.harvard.edu/abs/2017MNRAS.464.2174B} {464, 2174}

\bibitem[\protect\citeauthoryear{{Baumgardt} \& {Hilker}}{{Baumgardt} \&
  {Hilker}}{2018}]{Baumgardt2018}
{Baumgardt} H.,  {Hilker} M.,  2018, \mn@doi [\mnras] {10.1093/mnras/sty1057},
  \href {https://ui.adsabs.harvard.edu/abs/2018MNRAS.478.1520B} {478, 1520}

\bibitem[\protect\citeauthoryear{{Baumgardt} \& {Makino}}{{Baumgardt} \&
  {Makino}}{2003}]{Baumgardt2003}
{Baumgardt} H.,  {Makino} J.,  2003, \mn@doi [\mnras]
  {10.1046/j.1365-8711.2003.06286.x}, \href
  {https://ui.adsabs.harvard.edu/abs/2003MNRAS.340..227B} {340, 227}

\bibitem[\protect\citeauthoryear{{Baumgardt} \& {Vasiliev}}{{Baumgardt} \&
  {Vasiliev}}{2021}]{Baumgardt2021}
{Baumgardt} H.,  {Vasiliev} E.,  2021, \mn@doi [\mnras]
  {10.1093/mnras/stab1474}, \href
  {https://ui.adsabs.harvard.edu/abs/2021MNRAS.505.5957B} {505, 5957}

\bibitem[\protect\citeauthoryear{{Baumgardt}, {Makino}  \& {Hut}}{{Baumgardt}
  et~al.}{2005}]{Baumgardt2005}
{Baumgardt} H.,  {Makino} J.,   {Hut} P.,  2005, \mn@doi [\apj]
  {10.1086/426893}, \href
  {https://ui.adsabs.harvard.edu/abs/2005ApJ...620..238B} {620, 238}

\bibitem[\protect\citeauthoryear{{Baumgardt}, {Hilker}, {Sollima}  \&
  {Bellini}}{{Baumgardt} et~al.}{2019a}]{Baumgardt2019a}
{Baumgardt} H.,  {Hilker} M.,  {Sollima} A.,   {Bellini} A.,  2019a, \mn@doi
  [\mnras] {10.1093/mnras/sty2997}, \href
  {https://ui.adsabs.harvard.edu/abs/2019MNRAS.482.5138B} {482, 5138}

\bibitem[\protect\citeauthoryear{{Baumgardt} et~al.,}{{Baumgardt}
  et~al.}{2019b}]{Baumgardt2019b}
{Baumgardt} H.,  et~al., 2019b, \mn@doi [\mnras] {10.1093/mnras/stz2060}, \href
  {https://ui.adsabs.harvard.edu/abs/2019MNRAS.488.5340B} {488, 5340}

\bibitem[\protect\citeauthoryear{{Baumgardt}, {Sollima}  \&
  {Hilker}}{{Baumgardt} et~al.}{2020}]{Baumgardt2020}
{Baumgardt} H.,  {Sollima} A.,   {Hilker} M.,  2020, \mn@doi [\pasa]
  {10.1017/pasa.2020.38}, \href
  {https://ui.adsabs.harvard.edu/abs/2020PASA...37...46B} {37, e046}

\bibitem[\protect\citeauthoryear{{Bellazzini} et~al.,}{{Bellazzini}
  et~al.}{2008}]{Bellazzini2008}
{Bellazzini} M.,  et~al., 2008, \mn@doi [\aj] {10.1088/0004-6256/136/3/1147},
  \href {https://ui.adsabs.harvard.edu/abs/2008AJ....136.1147B} {136, 1147}

\bibitem[\protect\citeauthoryear{{Bianchini}, {Ibata}  \& {Famaey}}{{Bianchini}
  et~al.}{2019}]{Bianchini2019}
{Bianchini} P.,  {Ibata} R.,   {Famaey} B.,  2019, \mn@doi [\apjl]
  {10.3847/2041-8213/ab58d1}, \href
  {https://ui.adsabs.harvard.edu/abs/2019ApJ...887L..12B} {887, L12}

\bibitem[\protect\citeauthoryear{{Blandford}, {Meier}  \&
  {Readhead}}{{Blandford} et~al.}{2019}]{Blandford2019}
{Blandford} R.,  {Meier} D.,   {Readhead} A.,  2019, \mn@doi [\araa]
  {10.1146/annurev-astro-081817-051948}, \href
  {https://ui.adsabs.harvard.edu/abs/2019ARA&A..57..467B} {57, 467}

\bibitem[\protect\citeauthoryear{{Chernoff}, {Kochanek}  \&
  {Shapiro}}{{Chernoff} et~al.}{1986}]{Chernoff1986}
{Chernoff} D.~F.,  {Kochanek} C.~S.,   {Shapiro} S.~L.,  1986, \mn@doi [\apj]
  {10.1086/164591}, \href
  {https://ui.adsabs.harvard.edu/abs/1986ApJ...309..183C} {309, 183}

\bibitem[\protect\citeauthoryear{{Da Costa} \& {Freeman}}{{Da Costa} \&
  {Freeman}}{1976}]{DaCosta1976}
{Da Costa} G.~S.,  {Freeman} K.~C.,  1976, \mn@doi [\apj] {10.1086/154363},
  \href {https://ui.adsabs.harvard.edu/abs/1976ApJ...206..128D} {206, 128}

\bibitem[\protect\citeauthoryear{{Dalgleish} et~al.,}{{Dalgleish}
  et~al.}{2020}]{Dalgleish2020}
{Dalgleish} H.,  et~al., 2020, \mn@doi [\mnras] {10.1093/mnras/staa091}, \href
  {https://ui.adsabs.harvard.edu/abs/2020MNRAS.492.3859D} {492, 3859}

\bibitem[\protect\citeauthoryear{{Djorgovski} \& {King}}{{Djorgovski} \&
  {King}}{1986}]{Djorgovski1986}
{Djorgovski} S.,  {King} I.~R.,  1986, \mn@doi [\apjl] {10.1086/184685}, \href
  {https://ui.adsabs.harvard.edu/abs/1986ApJ...305L..61D} {305, L61}

\bibitem[\protect\citeauthoryear{{Ebisuzaki} et~al.,}{{Ebisuzaki}
  et~al.}{2001}]{Ebisuzaki2001}
{Ebisuzaki} T.,  et~al., 2001, \mn@doi [\apjl] {10.1086/338118}, \href
  {https://ui.adsabs.harvard.edu/abs/2001ApJ...562L..19E} {562, L19}

\bibitem[\protect\citeauthoryear{{Fitzpatrick}}{{Fitzpatrick}}{1999}]{Fitzpatrick1999}
{Fitzpatrick} E.~L.,  1999, \mn@doi [\pasp] {10.1086/316293}, \href
  {https://ui.adsabs.harvard.edu/abs/1999PASP..111...63F} {111, 63}

\bibitem[\protect\citeauthoryear{{Foreman-Mackey}, {Hogg}, {Lang}  \&
  {Goodman}}{{Foreman-Mackey} et~al.}{2013}]{Foreman-Mackey2013}
{Foreman-Mackey} D.,  {Hogg} D.~W.,  {Lang} D.,   {Goodman} J.,  2013, \mn@doi
  [\pasp] {10.1086/670067}, \href
  {https://ui.adsabs.harvard.edu/abs/2013PASP..125..306F} {125, 306}

\bibitem[\protect\citeauthoryear{{GRAVITY Collaboration} et~al.,}{{GRAVITY
  Collaboration} et~al.}{2019}]{GRAVITY2019}
{GRAVITY Collaboration} et~al., 2019, \mn@doi [\aap]
  {10.1051/0004-6361/201935656}, \href
  {https://ui.adsabs.harvard.edu/abs/2019A&A...625L..10G} {625, L10}

\bibitem[\protect\citeauthoryear{{Gerssen}, {van der Marel}, {Gebhardt},
  {Guhathakurta}, {Peterson}  \& {Pryor}}{{Gerssen} et~al.}{2002}]{Gerssen2002}
{Gerssen} J.,  {van der Marel} R.~P.,  {Gebhardt} K.,  {Guhathakurta} P.,
  {Peterson} R.~C.,   {Pryor} C.,  2002, \mn@doi [\aj] {10.1086/344584}, \href
  {https://ui.adsabs.harvard.edu/abs/2002AJ....124.3270G} {124, 3270}

\bibitem[\protect\citeauthoryear{{Gieles} \& {Zocchi}}{{Gieles} \&
  {Zocchi}}{2015}]{Gieles2015}
{Gieles} M.,  {Zocchi} A.,  2015, \mn@doi [\mnras] {10.1093/mnras/stv1848},
  \href {https://ui.adsabs.harvard.edu/abs/2015MNRAS.454..576G} {454, 576}

\bibitem[\protect\citeauthoryear{{Gill}, {Trenti}, {Miller}, {van der Marel},
  {Hamilton}  \& {Stiavelli}}{{Gill} et~al.}{2008}]{Gill2008}
{Gill} M.,  {Trenti} M.,  {Miller} M.~C.,  {van der Marel} R.,  {Hamilton} D.,
   {Stiavelli} M.,  2008, \mn@doi [\apj] {10.1086/591269}, \href
  {https://ui.adsabs.harvard.edu/abs/2008ApJ...686..303G} {686, 303}

\bibitem[\protect\citeauthoryear{{Gomez-Leyton} \& {Velazquez}}{{Gomez-Leyton}
  \& {Velazquez}}{2014}]{Gomez-Leyton2014}
{Gomez-Leyton} Y.~J.,  {Velazquez} L.,  2014, \mn@doi [Journal of Statistical
  Mechanics: Theory and Experiment] {10.1088/1742-5468/2014/04/P04006}, \href
  {https://ui.adsabs.harvard.edu/abs/2014JSMTE..04..006G} {2014, 04006}

\bibitem[\protect\citeauthoryear{{Goodman} \& {Weare}}{{Goodman} \&
  {Weare}}{2010}]{Goodman2010}
{Goodman} J.,  {Weare} J.,  2010, \mn@doi [Communications in Applied
  Mathematics and Computational Science] {10.2140/camcos.2010.5.65}, \href
  {https://ui.adsabs.harvard.edu/abs/2010CAMCS...5...65G} {5, 65}

\bibitem[\protect\citeauthoryear{{G{\"o}ttgens} et~al.,}{{G{\"o}ttgens}
  et~al.}{2021}]{Gottgens+2021}
{G{\"o}ttgens} F.,  et~al., 2021, \mn@doi [\mnras] {10.1093/mnras/stab2449},
  \href {https://ui.adsabs.harvard.edu/abs/2021MNRAS.507.4788G} {507, 4788}

\bibitem[\protect\citeauthoryear{{Gunn} \& {Griffin}}{{Gunn} \&
  {Griffin}}{1979}]{Gunn1979}
{Gunn} J.~E.,  {Griffin} R.~F.,  1979, \mn@doi [\aj] {10.1086/112477}, \href
  {https://ui.adsabs.harvard.edu/abs/1979AJ.....84..752G} {84, 752}

\bibitem[\protect\citeauthoryear{{H{\"a}berle} et~al.,}{{H{\"a}berle}
  et~al.}{2021}]{Haberle2021}
{H{\"a}berle} M.,  et~al., 2021, \mn@doi [\mnras] {10.1093/mnras/stab474},
  \href {https://ui.adsabs.harvard.edu/abs/2021MNRAS.503.1490H} {503, 1490}

\bibitem[\protect\citeauthoryear{{Harris}}{{Harris}}{1996}]{Harris1996}
{Harris} W.~E.,  1996, \mn@doi [\aj] {10.1086/118116}, \href
  {https://ui.adsabs.harvard.edu/abs/1996AJ....112.1487H} {112, 1487}

\bibitem[\protect\citeauthoryear{{Jindal}, {Webb}  \& {Bovy}}{{Jindal}
  et~al.}{2019}]{Jindal2019}
{Jindal} A.,  {Webb} J.~J.,   {Bovy} J.,  2019, \mn@doi [\mnras]
  {10.1093/mnras/stz1586}, \href
  {https://ui.adsabs.harvard.edu/abs/2019MNRAS.487.3693J} {487, 3693}

\bibitem[\protect\citeauthoryear{{Kamann} et~al.,}{{Kamann}
  et~al.}{2018}]{Kamann2018}
{Kamann} S.,  et~al., 2018, \mn@doi [\mnras] {10.1093/mnras/stx2719}, \href
  {https://ui.adsabs.harvard.edu/abs/2018MNRAS.473.5591K} {473, 5591}

\bibitem[\protect\citeauthoryear{{King}}{{King}}{1966}]{King1966}
{King} I.~R.,  1966, \mn@doi [\aj] {10.1086/109857}, \href
  {https://ui.adsabs.harvard.edu/abs/1966AJ.....71...64K} {71, 64}

\bibitem[\protect\citeauthoryear{{Lanzoni} et~al.,}{{Lanzoni}
  et~al.}{2013}]{Lanzoni2013}
{Lanzoni} B.,  et~al., 2013, \mn@doi [\apj] {10.1088/0004-637X/769/2/107},
  \href {https://ui.adsabs.harvard.edu/abs/2013ApJ...769..107L} {769, 107}

\bibitem[\protect\citeauthoryear{{Libralato} et~al.,}{{Libralato}
  et~al.}{2022}]{Libralato2022}
{Libralato} M.,  et~al., 2022, \mn@doi [\apj] {10.3847/1538-4357/ac7727}, \href
  {https://ui.adsabs.harvard.edu/abs/2022ApJ...934..150L} {934, 150}

\bibitem[\protect\citeauthoryear{{L{\"u}tzgendorf}, {Kissler-Patig}, {Noyola},
  {Jalali}, {de Zeeuw}, {Gebhardt}  \& {Baumgardt}}{{L{\"u}tzgendorf}
  et~al.}{2011}]{Lutzgendorf2011}
{L{\"u}tzgendorf} N.,  {Kissler-Patig} M.,  {Noyola} E.,  {Jalali} B.,  {de
  Zeeuw} P.~T.,  {Gebhardt} K.,   {Baumgardt} H.,  2011, \mn@doi [\aap]
  {10.1051/0004-6361/201116618}, \href
  {https://ui.adsabs.harvard.edu/abs/2011A&A...533A..36L} {533, A36}

\bibitem[\protect\citeauthoryear{{Lynden-Bell}}{{Lynden-Bell}}{1967}]{Lynden-Bell1967}
{Lynden-Bell} D.,  1967, \mn@doi [\mnras] {10.1093/mnras/136.1.101}, \href
  {https://ui.adsabs.harvard.edu/abs/1967MNRAS.136..101L} {136, 101}

\bibitem[\protect\citeauthoryear{{Manchester}, {Lyne}, {Robinson}, {D'Amico},
  {Bailes}  \& {Lim}}{{Manchester} et~al.}{1991}]{Manchester1991}
{Manchester} R.~N.,  {Lyne} A.~G.,  {Robinson} C.,  {D'Amico} N.,  {Bailes} M.,
    {Lim} J.,  1991, \mn@doi [\nat] {10.1038/352219a0}, \href
  {https://ui.adsabs.harvard.edu/abs/1991Natur.352..219M} {352, 219}

\bibitem[\protect\citeauthoryear{{McLaughlin} \& {van der Marel}}{{McLaughlin}
  \& {van der Marel}}{2005}]{McLaughlin2005}
{McLaughlin} D.~E.,  {van der Marel} R.~P.,  2005, \mn@doi [\apjs]
  {10.1086/497429}, \href
  {https://ui.adsabs.harvard.edu/abs/2005ApJS..161..304M} {161, 304}

\bibitem[\protect\citeauthoryear{{McLaughlin}, {Anderson}, {Meylan},
  {Gebhardt}, {Pryor}, {Minniti}  \& {Phinney}}{{McLaughlin}
  et~al.}{2006}]{McLaughlin2006}
{McLaughlin} D.~E.,  {Anderson} J.,  {Meylan} G.,  {Gebhardt} K.,  {Pryor} C.,
  {Minniti} D.,   {Phinney} S.,  2006, \mn@doi [\apjs] {10.1086/505692}, \href
  {https://ui.adsabs.harvard.edu/abs/2006ApJS..166..249M} {166, 249}

\bibitem[\protect\citeauthoryear{{McNamara}, {Harrison}  \&
  {Anderson}}{{McNamara} et~al.}{2003}]{McNamara2003}
{McNamara} B.~J.,  {Harrison} T.~E.,   {Anderson} J.,  2003, \mn@doi [\apj]
  {10.1086/377341}, \href
  {https://ui.adsabs.harvard.edu/abs/2003ApJ...595..187M} {595, 187}

\bibitem[\protect\citeauthoryear{{McNamara}, {Harrison}, {Baumgardt}  \&
  {Khalaj}}{{McNamara} et~al.}{2012}]{McNamara2012}
{McNamara} B.~J.,  {Harrison} T.~E.,  {Baumgardt} H.,   {Khalaj} P.,  2012,
  \mn@doi [\apj] {10.1088/0004-637X/745/2/175}, \href
  {https://ui.adsabs.harvard.edu/abs/2012ApJ...745..175M} {745, 175}

\bibitem[\protect\citeauthoryear{{Michie}}{{Michie}}{1963}]{Michie1963}
{Michie} R.~W.,  1963, \mn@doi [\mnras] {10.1093/mnras/125.2.127}, \href
  {https://ui.adsabs.harvard.edu/abs/1963MNRAS.125..127M} {125, 127}

\bibitem[\protect\citeauthoryear{{Mikolajewska}, {Zdziarski}, {Ziolkowski},
  {Torres}  \& {Casares}}{{Mikolajewska} et~al.}{2022}]{Mikolajewska2022}
{Mikolajewska} J.,  {Zdziarski} A.~A.,  {Ziolkowski} J.,  {Torres} M. A.~P.,
  {Casares} J.,  2022, \mn@doi [\apj] {10.3847/1538-4357/ac6099}, \href
  {https://ui.adsabs.harvard.edu/abs/2022ApJ...930....9M} {930, 9}

\bibitem[\protect\citeauthoryear{{Noyola}, {Gebhardt}, {Kissler-Patig},
  {L{\"u}tzgendorf}, {Jalali}, {de Zeeuw}  \& {Baumgardt}}{{Noyola}
  et~al.}{2010}]{Noyola2010}
{Noyola} E.,  {Gebhardt} K.,  {Kissler-Patig} M.,  {L{\"u}tzgendorf} N.,
  {Jalali} B.,  {de Zeeuw} P.~T.,   {Baumgardt} H.,  2010, \mn@doi [\apjl]
  {10.1088/2041-8205/719/1/L60}, \href
  {https://ui.adsabs.harvard.edu/abs/2010ApJ...719L..60N} {719, L60}

\bibitem[\protect\citeauthoryear{{Oh} \& {Lin}}{{Oh} \& {Lin}}{1992}]{Oh1992}
{Oh} K.~S.,  {Lin} D.~N.~C.,  1992, \mn@doi [\apj] {10.1086/171037}, \href
  {https://ui.adsabs.harvard.edu/abs/1992ApJ...386..519O} {386, 519}

\bibitem[\protect\citeauthoryear{{Oort} \& {van Herk}}{{Oort} \& {van
  Herk}}{1959}]{Oort1959}
{Oort} J.~H.,  {van Herk} G.,  1959, \bain, \href
  {https://ui.adsabs.harvard.edu/abs/1959BAN....14..299O} {14, 299}

\bibitem[\protect\citeauthoryear{{Peuten}, {Zocchi}, {Gieles}, {Gualandris}  \&
  {H{\'e}nault-Brunet}}{{Peuten} et~al.}{2016}]{Peuten2016}
{Peuten} M.,  {Zocchi} A.,  {Gieles} M.,  {Gualandris} A.,
  {H{\'e}nault-Brunet} V.,  2016, \mn@doi [\mnras] {10.1093/mnras/stw1726},
  \href {https://ui.adsabs.harvard.edu/abs/2016MNRAS.462.2333P} {462, 2333}

\bibitem[\protect\citeauthoryear{{Peuten}, {Zocchi}, {Gieles}  \&
  {H{\'e}nault-Brunet}}{{Peuten} et~al.}{2017}]{Peuten2017}
{Peuten} M.,  {Zocchi} A.,  {Gieles} M.,   {H{\'e}nault-Brunet} V.,  2017,
  \mn@doi [\mnras] {10.1093/mnras/stx1311}, \href
  {https://ui.adsabs.harvard.edu/abs/2017MNRAS.470.2736P} {470, 2736}

\bibitem[\protect\citeauthoryear{{Plummer}}{{Plummer}}{1911}]{Plummer1911}
{Plummer} H.~C.,  1911, \mn@doi [\mnras] {10.1093/mnras/71.5.460}, \href
  {https://ui.adsabs.harvard.edu/abs/1911MNRAS..71..460P} {71, 460}

\bibitem[\protect\citeauthoryear{{Sanna}, {Pancino}, {Zocchi}, {Ferraro}  \&
  {Stetson}}{{Sanna} et~al.}{2020}]{Sanna+2020}
{Sanna} N.,  {Pancino} E.,  {Zocchi} A.,  {Ferraro} F.~R.,   {Stetson} P.~B.,
  2020, \mn@doi [\aap] {10.1051/0004-6361/202037500}, \href
  {https://ui.adsabs.harvard.edu/abs/2020A&A...637A..46S} {637, A46}

\bibitem[\protect\citeauthoryear{{Spitzer} \& {Shapiro}}{{Spitzer} \&
  {Shapiro}}{1972}]{Spitzer1972}
{Spitzer} Lyman J.,  {Shapiro} S.~L.,  1972, \mn@doi [\apj] {10.1086/151442},
  \href {https://ui.adsabs.harvard.edu/abs/1972ApJ...173..529S} {173, 529}

\bibitem[\protect\citeauthoryear{{Takahashi}, {Lee}  \& {Inagaki}}{{Takahashi}
  et~al.}{1997}]{Takahashi1997}
{Takahashi} K.,  {Lee} H.~M.,   {Inagaki} S.,  1997, \mn@doi [\mnras]
  {10.1093/mnras/292.2.331}, \href
  {https://ui.adsabs.harvard.edu/abs/1997MNRAS.292..331T} {292, 331}

\bibitem[\protect\citeauthoryear{{Tiongco}, {Vesperini}  \& {Varri}}{{Tiongco}
  et~al.}{2016}]{Tiongco2016}
{Tiongco} M.~A.,  {Vesperini} E.,   {Varri} A.~L.,  2016, \mn@doi [\mnras]
  {10.1093/mnras/stv2574}, \href
  {https://ui.adsabs.harvard.edu/abs/2016MNRAS.455.3693T} {455, 3693}

\bibitem[\protect\citeauthoryear{{Trager}, {King}  \& {Djorgovski}}{{Trager}
  et~al.}{1995}]{Trager1995}
{Trager} S.~C.,  {King} I.~R.,   {Djorgovski} S.,  1995, \mn@doi [\aj]
  {10.1086/117268}, \href
  {https://ui.adsabs.harvard.edu/abs/1995AJ....109..218T} {109, 218}

\bibitem[\protect\citeauthoryear{{Vandenberg}, {Bolte}  \&
  {Stetson}}{{Vandenberg} et~al.}{1996}]{Vandenberg1996}
{Vandenberg} D.~A.,  {Bolte} M.,   {Stetson} P.~B.,  1996, \mn@doi [\araa]
  {10.1146/annurev.astro.34.1.461}, \href
  {https://ui.adsabs.harvard.edu/abs/1996ARA&A..34..461V} {34, 461}

\bibitem[\protect\citeauthoryear{{Vasiliev} \& {Baumgardt}}{{Vasiliev} \&
  {Baumgardt}}{2021}]{Vasiliev2021}
{Vasiliev} E.,  {Baumgardt} H.,  2021, \mn@doi [\mnras]
  {10.1093/mnras/stab1475}, \href
  {https://ui.adsabs.harvard.edu/abs/2021MNRAS.505.5978V} {505, 5978}

\bibitem[\protect\citeauthoryear{{Wan} et~al.,}{{Wan} et~al.}{2021}]{Wan2021}
{Wan} Z.,  et~al., 2021, \mn@doi [\mnras] {10.1093/mnras/stab306}, \href
  {https://ui.adsabs.harvard.edu/abs/2021MNRAS.502.4513W} {502, 4513}

\bibitem[\protect\citeauthoryear{{Watkins}, {van der Marel}, {Bellini}  \&
  {Anderson}}{{Watkins} et~al.}{2015a}]{Watkins2015a}
{Watkins} L.~L.,  {van der Marel} R.~P.,  {Bellini} A.,   {Anderson} J.,
  2015a, \mn@doi [\apj] {10.1088/0004-637X/803/1/29}, \href
  {https://ui.adsabs.harvard.edu/abs/2015ApJ...803...29W} {803, 29}

\bibitem[\protect\citeauthoryear{{Watkins}, {van der Marel}, {Bellini}  \&
  {Anderson}}{{Watkins} et~al.}{2015b}]{Watkins2015b}
{Watkins} L.~L.,  {van der Marel} R.~P.,  {Bellini} A.,   {Anderson} J.,
  2015b, \mn@doi [\apj] {10.1088/0004-637X/812/2/149}, \href
  {https://ui.adsabs.harvard.edu/abs/2015ApJ...812..149W} {812, 149}

\bibitem[\protect\citeauthoryear{{Wilson}}{{Wilson}}{1975}]{Wilson1975}
{Wilson} C.~P.,  1975, \mn@doi [\aj] {10.1086/111729}, \href
  {https://ui.adsabs.harvard.edu/abs/1975AJ.....80..175W} {80, 175}

\bibitem[\protect\citeauthoryear{{Woolley}}{{Woolley}}{1954}]{Woolley1954}
{Woolley} R.~V.~D.~R.,  1954, \mn@doi [\mnras] {10.1093/mnras/114.2.191}, \href
  {https://ui.adsabs.harvard.edu/abs/1954MNRAS.114..191W} {114, 191}

\bibitem[\protect\citeauthoryear{{Zloczewski}, {Kaluzny}, {Rozyczka},
  {Krzeminski}  \& {Mazur}}{{Zloczewski} et~al.}{2012}]{Zloczewski2012}
{Zloczewski} K.,  {Kaluzny} J.,  {Rozyczka} M.,  {Krzeminski} W.,   {Mazur} B.,
   2012, \actaa, \href {https://ui.adsabs.harvard.edu/abs/2012AcA....62..357Z}
  {62, 357}

\bibitem[\protect\citeauthoryear{{Zocchi}, {Gieles}, {H{\'e}nault-Brunet}  \&
  {Varri}}{{Zocchi} et~al.}{2016}]{Zocchi2016}
{Zocchi} A.,  {Gieles} M.,  {H{\'e}nault-Brunet} V.,   {Varri} A.~L.,  2016,
  \mn@doi [\mnras] {10.1093/mnras/stw1104}, \href
  {https://ui.adsabs.harvard.edu/abs/2016MNRAS.462..696Z} {462, 696}

\bibitem[\protect\citeauthoryear{{Zocchi}, {Gieles}  \&
  {H{\'e}nault-Brunet}}{{Zocchi} et~al.}{2017}]{Zocchi2017}
{Zocchi} A.,  {Gieles} M.,   {H{\'e}nault-Brunet} V.,  2017, \mn@doi [\mnras]
  {10.1093/mnras/stx316}, \href
  {https://ui.adsabs.harvard.edu/abs/2017MNRAS.468.4429Z} {468, 4429}

\bibitem[\protect\citeauthoryear{{den Brok}, {van de Ven}, {van den Bosch}  \&
  {Watkins}}{{den Brok} et~al.}{2014}]{denBrok2014}
{den Brok} M.,  {van de Ven} G.,  {van den Bosch} R.,   {Watkins} L.,  2014,
  \mn@doi [\mnras] {10.1093/mnras/stt2221}, \href
  {https://ui.adsabs.harvard.edu/abs/2014MNRAS.438..487D} {438, 487}

\bibitem[\protect\citeauthoryear{{van Leeuwen}}{{van
  Leeuwen}}{2009}]{vanLeeuwen2009}
{van Leeuwen} F.,  2009, \mn@doi [\aap] {10.1051/0004-6361/200811382}, \href
  {https://ui.adsabs.harvard.edu/abs/2009A&A...497..209V} {497, 209}

\makeatother
\end{thebibliography}



\bsp	
\label{lastpage}
\end{document}